\documentclass[12pt]{article} 
\pdfoutput=1
\usepackage{graphicx,floatflt,amssymb,rotate}
\usepackage{mathrsfs}
\usepackage{amssymb}
\usepackage{amsmath}
\usepackage{color}
\usepackage{graphicx}
\usepackage{subfigure}
\usepackage{multirow}
\usepackage{float}
\usepackage{pstricks}
\usepackage[toc,page]{appendix}
\usepackage{cite}
\usepackage[colorlinks=true,
linkcolor=red,
urlcolor=blue,
citecolor=blue]{hyperref}
\usepackage{color}

\newcommand{\RE}{{\rm Re}}

\def\R1{\varepsilon_1}
\def\E8{\varepsilon_8}

\newcommand{\bea}{\begin{eqnarray}}
\newcommand{\eea}{\end{eqnarray}}
\newcommand{\bd}{\begin{displaymath}}
\newcommand{\ed}{\end{displaymath}}

\newcommand{\be}{\begin{equation}}
\newcommand{\ee}{\end{equation}}
\newcommand{\bi}{\begin{itemize}}
\newcommand{\ei}{\end{itemize}}

\newcommand{\Ctilde}{\tilde{C}}

\begin{document}
\vskip 30pt  
 
\begin{center}  
{\Large \bf Looking for \boldmath$B\rightarrow X_s \ell^+\ell^-$ in non-minimal Universal Extra Dimensional model} \\
\vspace*{1cm}  
\renewcommand{\thefootnote}{\fnsymbol{footnote}}  
{{\sf Avirup Shaw$^1$\footnote{email: avirup.cu@gmail.com}} 
}\\  
\vspace{10pt}  
{{\em $^1$Theoretical Physics, Physical Research Laboratory,\\
Ahmedabad 380009, India}}
\normalsize  
\end{center} 

\begin{abstract}
\noindent 
Non-vanishing boundary localised terms significantly modify the mass spectrum and various interactions among the Kaluza-Klein excited states of 5-Dimensional Universal Extra Dimensional scenario. In this scenario we compute the contributions of Kaluza-Klein excitations of gauge bosons and third generation quarks for the decay process $B\rightarrow X_s\ell^+\ell^-$ incorporating next-to-leading order QCD corrections. We estimate branching ratio as well as Forward Backward asymmetry associated with this decay process. Considering the constraints from some other $b \to s$ observables and electroweak precision data we show that significant amount of parameter space of this scenario has been able to explain the observed experimental data for this decay process. From our analysis we put lower limit on the size of the extra dimension by comparing our theoretical prediction for branching ratio with the corresponding experimental data. Depending on the values of free parameters of the present scenario, lower limit on the inverse of the radius of compactification ($R^{-1}$) can be as high as $\geq 760$ GeV. {Even this value could slightly be higher if we project the upcoming measurement by Belle II experiment.} Unfortunately, the Forward Backward asymmetry of this decay process would not provide any significant limit on $R^{-1}$ in the present model.
\vskip 5pt \noindent 
\end{abstract}

%\noindent PACS No: {\tt 11.10.Kk, 12.60.-i, 13.20.He}\\
%\texttt{Key Words:~~Universal Extra Dimension, Kaluza-Klein, radiative decays of mesons} 
\renewcommand{\thesection}{\Roman{section}}  
\setcounter{footnote}{0}  
\renewcommand{\thefootnote}{\arabic{footnote}}

\section{Introduction}
Standard Model (SM) of particle physics has {\it almost} been accomplished by the discovery of Higgs boson at the Large Hadron Collider (LHC) \cite{Aad:2012tfa, Chatrchyan:2012xdj}. However, the SM scenario is not the ultimate one, because there exists several experimental data in various directions, such as massive neutrinos, Dark Matter (DM) enigma, observed baryon asymmetry etc., that cannot be addressed within the SM. This in turn, ensures that new physics (NP) is indeed a reality of nature. Moreover, experimental data for several flavour (specially $B$-physics) physics observables show significant deviation from the corresponding SM expectations. For example, in $B$-physics experiments at LHCb, Belle and Babar have pointed at intriguing lepton flavour universality violating (LFUV) effects for both the charge current ($\mathcal{R}_{D^{(*)}}$~\cite{average} and $\mathcal{R}_{J/\psi}$~\cite{Aaij:2017tyk}) as well as the flavour changing neutral current (FCNC) ($\mathcal{R}_K$ \cite{Aaij:2019wad} and $\mathcal{R}_{K^*}$\cite{Aaij:2017vbb}) processes. In the latter case, involving processes are described at the quark level by the transition $b \to s \ell^+ \ell^- (\ell \equiv e-{\rm the\;electron}, \mu-{\rm the\;muon})$  which is highly suppressed in SM. Therefore, even for small deviation between the SM prediction and experimental data, these types of observables have always been very instrumental to probe the favorable kind among the various NP models that exist in the literature. {Apart from these, there exists several $B$-physics observables which could also be used for the detection of NP scenarios.} 

{Following the above argument in the current article, we will calculate an inclusive decay mode $B\rightarrow X_s \ell^+\ell^-$ in a NP scenario namely non-minimal Universal Extra Dimensional (nmUED) model\footnote{In this model we have already calculated several $B$-physics obeservables, for example branching fractions of some rare decay processes e.g., $B_s \rightarrow \mu^+ \mu^-$ \cite{Datta:2015aka}, $B \to X_s\gamma$ \cite{Datta:2016flx} and $\mathcal{R}_{D^{(*)}}$ anomalies \cite{Biswas:2017vhc}.}. This inclusive decay mode $B\rightarrow X_s \ell^+\ell^-$ has been considered as one of the harbingers for the detection of NP scenario. The reason is that, this decay mode is one of the most significant and relatively clean decay modes. $B\to X_s\ell^+\ell^-$ decay is significant in a sense that this decay mode not only helps the detection of NP scenario but also presents more complex test of the SM. For example, in comparison with the $B\to X_s\gamma$ decay, different contributions add to the inclusive $B\to X_s\ell^+\ell^-$ decay. Moreover, it is particularly attractive because, as a three body decay process it also offers more kinematic observables such as the invariant di-lepton mass spectrum and the Forward-Backward asymmetry \cite{Hurth:2003vb, Benzke:2017woq}. At the quark level this process is also governed by $b \to s \ell^+ \ell^-$ transition. The effective Hamiltonian of this decay process is characterised by three different Wilson Coefficients (WCs): $C_7$, $C_9$ and $C_{10}$. Among these WCs, $C_{10}$ and $C_7$ for nmUED model have already been calculated in our previous studies \cite{Datta:2015aka} and \cite{Datta:2016flx} respectively. Consequently, calculation of the WC $C_9$ using relevant one loop Feynman diagrams in the context of nmUED model is one of the primary tasks of this article. The full calculational details of the WC $C_9$ have been given in Sec.\;\ref{sec:Heff:BXsee:nlo}. To the best of our knowledge, this is to be the first article where we will show the calculation of the WC $C_9$ in the context of nmUED model in detail. Finally, with these different WCs $C_7$, $C_9$ and $C_{10}$ we compute the coefficients of electroweak dipole operators for photon and gluon for the first time in the nmUED scenario. Eventually, we can readily calculate the decay amplitude for this process $B\rightarrow X_s \ell^+\ell^-$ in the nmUED scenario.}

In most of the cases, experimental data for several observables for the decay mode $B\rightarrow X_s \ell^+\ell^-$ have been more explored for two regions\footnote{The reason for choosing these two regions is given in Sec.\;\ref{sec:Heff:BXsee:nlo}.} of di-lepton invariant mass square $q^2$ \bigg($\equiv (p_{_{\ell^+}}+p_{_{\ell^-}})^2$\bigg) spectrum. In these two regions, the experimental data of branching ratio (Br) are given by Babar collaboration\footnote{These experimental data have also been used in two recent articles \cite{Feng:2016wph, Kumar:2019qbv} in context of the same decay process.} \cite{Lees:2013nxa}
\begin{eqnarray}
%%%%%%%%%%%%%%%%%%%%%%%%%%%%%%%%%%%%%%%%%%%%%%%%%%%
&&{\rm Br}(B\rightarrow X_{_s}\ell^+\ell^-)_{_{q^2\in[1,6]{\rm GeV}^2}}^{\rm exp}=(1.60^{+0.41+0.17}_{-0.39-0.13}\pm 0.18)\times10^{-6},\;
\nonumber\\
&&{\rm Br}(B\rightarrow X_{_s}\ell^+\ell^-)_{_{q^2\in[14.4,\;25]{\rm GeV}^2}}^{\rm exp}=(0.57^{+0.16+0.03}_{-0.15-0.02}\pm 0.00)\times10^{-6}
\;,(\ell=e,\;\mu)\;.
%%%%%%%%%%%%%%%%%%%%%%%%%%%%%%%%%%%%%%%%%%%%%%%%%%%
\label{EXP-BR-BtoXsll}
\end{eqnarray} 
The SM predictions for the above quantities are \cite{Huber:2015sra}
\begin{eqnarray}
%%%%%%%%%%%%%%%%%%%%%%%%%%%%%%%%%%%%%%%%%%%%%%%%%%%
&&{\rm Br}(B\rightarrow X_{_s}\ell^+\ell^-)_{_{q^2\in[1,6]{\rm GeV}^2}}^{\rm SM}=(1.62\pm 0.09)\times10^{-6},\;
\nonumber\\
&&{\rm Br}(B\rightarrow X_{_s}\ell^+\ell^-)_{_{q^2\in[14.4,25]{\rm GeV}^2}}^{\rm SM}=(2.53\pm 0.70)\times10^{-7}\;,
(\ell=e,\;\mu)\;.
%%%%%%%%%%%%%%%%%%%%%%%%%%%%%%%%%%%%%%%%%%%%%%%%%%%
\label{SM-BR-BtoXsll}
\end{eqnarray}
Moreover, apart from the branching ratio, Forward-Backward asymmetry ($A_{FB}$) could also help for the detection of NP  scenario. For this decay process $B\rightarrow X_{_S}\ell^+\ell^-\;(\ell=e,\;\mu)$ for the two distinct regions of $q^2$ the experimental values of this observable are given by Belle Collaboration\cite{Sato:2014pjr}
\begin{eqnarray}
%%%%%%%%%%%%%%%%%%%%%%%%%%%%%%%%%%%%%%%%%%%%%%%%%%%
&&A_{_{FB}}(B\rightarrow X_{_s}\ell^+\ell^-)\Big|_{q^2\in[1,\;6]\;{\rm GeV}^2}^{\rm exp}
=0.30\pm0.24\pm0.04\;,
\nonumber\\
&&A_{_{FB}}(B\rightarrow X_{_s}\ell^+\ell^-)\Big|_{q^2\in[14.4,\;25]\;{\rm GeV}^2}^{\rm exp}
=0.28\pm0.15\pm0.02\;,
%%%%%%%%%%%%%%%%%%%%%%%%%%%%%%%%%%%%%%%%%%%%%%%%%%%
\label{exp-AFB0}
\end{eqnarray}
while the corresponding SM expectations are \cite{Fukae:1998qy, Ali:2002jg, Sato:2014pjr}
\begin{eqnarray}
%%%%%%%%%%%%%%%%%%%%%%%%%%%%%%%%%%%%%%%%%%%%%%%%%%%
&&A_{_{FB}}(B\rightarrow X_{_s}\ell^+\ell^-)\Big|_{q^2\in[1,\;6]\;{\rm GeV}^2}^{\rm SM}
=-0.07\pm0.04\;,
\nonumber\\
&&A_{_{FB}}(B\rightarrow X_{_s}\ell^+\ell^-)\Big|_{q^2\in[14.4,\;25]\;{\rm GeV}^2}^{\rm SM}
=0.40\pm0.04\;.
%%%%%%%%%%%%%%%%%%%%%%%%%%%%%%%%%%%%%%%%%%%%%%%%%%%
\label{SM-AFB0}
\end{eqnarray}
Therefore, from the above data it is clearly evident that the SM predictions for the respective observables coincide with the experimental data within a few standard deviations. Hence, by investigating these observables one can search any favourable NP scenario and also tightly constrain the parameter space of that scenario. With this spirit, in this article we evaluate the decay amplitude for the process $B\rightarrow X_{_s}\ell^+\ell^-$ in nmUED scenario. In literature one can find several articles, e.g., \cite{Feng:2016wph, Lunghi:1999uk}  which have been dedicated to the exploration of the same decay process in the context of several beyond SM (BSM) scenarios.

In the present article, {in order to serve our purposes} we are particularly focused on an extension of SM with one flat space-like dimension ($y$) compactified on a circle $S^1$ of radius $R$. All the SM fields are allowed to propagate along the extra dimension $y$. This model is called as 5-dimensional (5D) Universal Extra Dimensional (UED) \cite{Appelquist:2000nn} scenario. The fields manifested on this manifold are usually defined in terms of towers of 4-Dimensional (4D) Kaluza-Klein (KK) states while the zero-mode of the KK-towers is designated as the corresponding 4D SM field. A discrete symmetry ${Z}_2$ ($y \leftrightarrow -y$) has been needed to generate chiral SM fermions in this scenario. Consequently, the extra dimension is defined as $S^1/Z_2$ orbifold and eventually physical domain extends from $y = 0$ to $y = \pi R$. As a result, the $y \leftrightarrow -y$ symmetry has been translated as a conserved parity which is known as KK-parity  $=(-1)^n$, where $n$ is called the KK-number. This KK-number ($n$) is identified as discretised momentum along the $y$-direction. From the conservation of KK-parity the lightest Kaluza-Klein particle (LKP) with KK-number one ($n=1$) cannot decay to a pair of SM particles and becomes absolutely stable. Hence, the LKP has been considered as a potential DM candidate in this scenario \cite{Servant:2002hb, Servant:2002aq, Cheng:2002ej, Majumdar:2002mw, Burnell:2005hm, Kong:2005hn, Kakizaki:2006dz, Belanger:2010yx}. Furthermore,  few variants of this model can address some other shortcomings of SM, for example, gauge coupling unifications \cite{Dienes:1998vh, Dienes:1998vg, Bhattacharyya:2006ym}, neutrino mass \cite{Hsieh:2006qe, Fujimoto:2014fka} and fermion mass hierarchy \cite{Archer:2012qa} etc. 

At the $n^{th}$ KK-level all the KK-state particles have the mass $\sqrt{(m^2+(nR^{-1})^2)}$. Here, $m$ is considered as the zero-mode mass (SM particle mass) which is very small with respect to $R^{-1}$. Therefore, this UED scenario contains almost degenerate mass spectrum at each KK-level. Consequently, this scenario has lost its phenomenological relevance, specifically, at the colliders. However, this degeneracy in the mass spectrum can be lifted by radiative corrections \cite{Georgi:2000ks, Cheng:2002iz}. There are two different types of radiative corrections. The first one is considered as bulk corrections (which are finite and only non-zero for KK-excitations of gauge bosons) and second one is regarded as boundary localised corrections that proportional to logarithmically cut-off\footnote{UED is considered as an effective theory and it is characterised by a cut-off scale $\Lambda$.} scale ($\Lambda$) dependent terms. The boundary correction terms can be embedded as 4D kinetic, mass and other possible interaction terms for the KK-states at the two fixed boundary points ($y=0$ and $y=\pi R$) of this orbifold. As a matter of fact, it is very obvious to include such terms in an extra dimensional theory like UED since these boundary terms have played the role of counterterms for cut-off dependent loop-induced contributions. In the minimal version of UED (mUED) models there is an assumption that these boundary terms are tuned in such a way that the 5D radiative corrections exactly vanish at the cut-off scale $\Lambda$. However, in general this assumption can be avoided and without calculating the actual radiative corrections one might consider kinetic, mass as well as other interaction terms localised at the two fixed boundary points to parametrise these unknown corrections. Therefore, this specific scenario is called as nmUED \cite{Dvali:2001gm, Carena:2002me, delAguila:2003bh, delAguila:2003kd, delAguila:2003gv, Schwinn:2004xa, Flacke:2008ne, Datta:2012xy, Flacke:2013pla}. In this scenario, not only the radius of compactification ($R$), but also the coefficients of different boundary localised terms (BLTs) have been considered as free parameters which can be  constrained by various experimental data of different physical observables. In literature one can find different such exercise regarding various phenomenological aspects. As for example limits on the values of the strengths of the BLTs have been achieved from the estimation of electroweak observables \cite{Flacke:2008ne, Flacke:2013pla}, S, T and U parameters \cite{delAguila:2003gv, Flacke:2013nta}, DM relic density \cite{Bonnevier:2011km, Datta:2013nua}, production as well as decay of SM Higgs boson \cite{Dey:2013cqa}, collider study of LHC experiments \cite{Datta:2012tv, Datta:2013yaa, Datta:2013lja, Shaw:2014gba, Shaw:2017whr, Ganguly:2018pzs}, $R_b$ \cite{Jha:2014faa}, branching ratios of some rare decay processes e.g., $B_s \rightarrow \mu^+ \mu^-$ \cite{Datta:2015aka} and $B \to X_s\gamma$ \cite{Datta:2016flx}, $\mathcal{R}_{D^{(*)}}$ anomalies \cite{Biswas:2017vhc, Dasgupta:2018nzt}, flavour changing rare top decay \cite{Dey:2016cve, Chiang:2018oyd} and unitarity of scattering amplitudes involving KK-excitations \cite{Jha:2016sre}.

In this article we estimate the contributions of KK-excited modes to the decay of $B\rightarrow X_s\ell^+\ell^-$ in a 5D UED model with {\it non-vanishing} BLT parameters. Our calculation includes next-to-leading order (NLO) QCD corrections. To the best of our knowledge, this is to be the first article where we will study the decay of $B\rightarrow X_s\ell^+\ell^-$ in the framework of nmUED.  Considering the present experimental data of the concerned FCNC process we will put constraints on the BLT parameters. Furthermore, we would like to investigate how far the lower limit on $R^{-1}$ to higher values can be extended using non-zero BLT parameters. Consequently, it will be an interesting part of this exercise is to see whether this lower limit of $R^{-1}$ is comparable with the results obtained from our previous analysis \cite{Datta:2015aka, Datta:2016flx} or not? Several years ago the same analysis \cite{Buras:2003mk} has been performed in the context of minimal version of UED model, however, the present experimental data have been changed with respect to that time. Therefore, it will be a relevant job to revisit the lower bound on {$R^{-1}$} in UED model by comparing the current experimental result \cite{Lees:2013nxa, Sato:2014pjr} with theoretical estimation using {\it vanishing} BLT parameters. {Furthermore, we estimate the probable bounds on the parameter space of the nmUED scenario by considering the upcoming measurement by Belle II experiment for the $B\rightarrow X_s\ell^+\ell^-$ decay observables.}

In the following section \ref{model}, we will give a brief description of the nmUED model. Then in section \ref{sec:Heff:BXsee:nlo} we will show the calculational details of branching ratio and Forward-Backward asymmetry for the present process. In section \ref{anls} we will present our numerical results. Finally, we conclude the results in section \ref{concl}.

\section{KK-parity conserving nmUED scenario: A brief overview}\label{model}
Here we present the technicalities of nmUED scenario required for our analysis. For further discussion regarding this scenario one can look into\cite{Dvali:2001gm, Carena:2002me, delAguila:2003bh, delAguila:2003kd, delAguila:2003gv, Schwinn:2004xa, Flacke:2008ne, Datta:2012xy, Datta:2012tv, Datta:2013yaa, Datta:2013lja, Shaw:2014gba, Shaw:2017whr, Ganguly:2018pzs, Jha:2014faa, Datta:2015aka, Datta:2016flx, Biswas:2017vhc}.
In the present scenario we preserve a $Z_2$ symmetry by considering equal strength of boundary terms at both the boundary points ($y=0$ and $y=\pi R$). Consequently, KK-parity has been restored in this scenario which makes the LKP stable. Hence, this present scenario can give a potential DM candidate (such as first excited KK-state of photon). A comprehensive exercise on DM in nmUED can be found in \cite{Datta:2013nua}.
 
We begin with the action for 5D fermionic fields associated with their boundary localised kinetic term (BLKT) of strength $r_f$ \cite{Schwinn:2004xa, Datta:2013nua, Datta:2015aka, Datta:2016flx, Biswas:2017vhc}
\begin{eqnarray} 
S_{fermion} = \int d^5x \left[ \bar{\Psi}_L i \Gamma^M D_M \Psi_L 
+ r_f\{\delta(y)+\delta(y - \pi R)\} \bar{\Psi}_L i \gamma^\mu D_\mu P_L\Psi_L  
\right. \nonumber \\
\left. + \bar{\Psi}_R i \Gamma^M D_M \Psi_R
+ r_f\{\delta(y)+\delta(y - \pi R)\}\bar{\Psi}_R i \gamma^\mu D_\mu P_R\Psi_R
\right],
\label{factn}
\end{eqnarray}
where $\Psi_L(x,y)$ and $\Psi_R(x,y)$ represent the 5D four component Dirac spinors that can be expressed in terms of two component spinors as \cite{Schwinn:2004xa, Datta:2013nua, Datta:2015aka, Datta:2016flx, Biswas:2017vhc}
\begin{equation} 
\Psi_L(x,y) = \begin{pmatrix}\phi_L(x,y) \\ \chi_L(x,y)\end{pmatrix}
=   \sum_n \begin{pmatrix}\phi^{(n)}_L(x) f_L^n(y) \\ \chi^{(n)}_L(x) g_L^n(y)\end{pmatrix}, 
\label{fermionexpnsn1}
\end{equation}
\begin{equation} 
\Psi_R(x,y) = \begin{pmatrix}\phi_R(x,y) \\ \chi_R(x,y) \end{pmatrix} 
=   \sum_n \begin{pmatrix}\phi^{(n)}_R(x) f_R^n(y) \\ \chi^{(n)}_R(x) g_R^n(y) \end{pmatrix}. 
\label{fermionexpnsn2} 
\end{equation}\\
$f_{L(R)}$ and $g_{L(R)}$ are the associated KK-wave-functions which can be written as the following \cite{Carena:2002me, Flacke:2008ne, Datta:2013nua, Datta:2015aka, Datta:2016flx, Biswas:2017vhc}
\begin{eqnarray}
f_L^n = g_R^n = N^f_n \left\{ \begin{array}{rl}
                \displaystyle \frac{\cos\left[m_{f^{(n)}} \left (y - \frac{\pi R}{2}\right)\right]}{\cos[ \frac{m_{f^{(n)}} \pi R}{2}]}  &\mbox{for $n$ even,}\\
                \displaystyle \frac{{-}\sin\left[m_{f^{(n)}} \left (y - \frac{\pi R}{2}\right)\right]}{\sin[ \frac{m_{f^{(n)}} \pi R}{2}]} &\mbox{for $n$ odd,}
                \end{array} \right.
                \label{flgr}
\end{eqnarray}
and
\begin{eqnarray}
g_L^n =-f_R^n = N^f_n \left\{ \begin{array}{rl}
                \displaystyle \frac{\sin\left[m_{f^{(n)}} \left (y - \frac{\pi R}{2}\right)\right]}{\cos[ \frac{m_{f^{(n)}} \pi R}{2}]}  &\mbox{for $n$ even,}\\
                \displaystyle \frac{\cos\left[m_{f^{(n)}} \left (y - \frac{\pi R}{2}\right)\right]}{\sin[ \frac{m_{f^{(n)}} \pi R}{2}]} &\mbox{for $n$ odd.}
                \end{array} \right.
\end{eqnarray}
Normalisation constant ($N^f_n$) for $n^{th}$ KK-mode can easily be obtained from the following orthonormality conditions \cite{Datta:2013nua, Datta:2015aka, Datta:2016flx, Biswas:2017vhc}
\begin{equation}\label{orthonorm}
\begin{aligned}
&\left.\begin{array}{r}
                  \int_0 ^{\pi R}
dy \; \left[1 + r_{f}\{ \delta(y) + \delta(y - \pi R)\}\right]f_L^mf_L^n\\
                  \int_0 ^{\pi R}
dy \; \left[1 + r_{f}\{ \delta(y) + \delta(y - \pi R)\}\right]g_R^mg_R^n
\end{array}\right\}=&&\delta^{n m}~;
&&\left.\begin{array}{l}
                 \int_0 ^{\pi R}
dy \; f_R^mf_R^n\\
                 \int_0 ^{\pi R}
dy \; g_L^mg_L^n
\end{array}\right\}=&&\delta^{n m}~,
\end{aligned}
\end{equation}
and it takes the form as
{\small
\vspace*{-.3cm}
\begin{equation}\label{norm}
N^f_n=\sqrt{\frac{2}{\pi R}}\Bigg[ \frac{1}{\sqrt{1 + \frac{r^2_f m^2_{f^{(n)}}}{4} + \frac{r_f}{\pi R}}}\Bigg].
\end{equation}
}

Here, $m_{f^{(n)}}$ is the KK-mass of $n^{th}$ KK-excitation acquired from the given transcendental equations \cite{Carena:2002me, Datta:2013nua, Datta:2015aka, Datta:2016flx, Biswas:2017vhc} 
\begin{eqnarray}
  \frac{r_{f} m_{f^{(n)}}}{2}= \left\{ \begin{array}{rl}
         -\tan \left(\frac{m_{f^{(n)}}\pi R}{2}\right) &\mbox{for $n$ even,}\\
          \cot \left(\frac{m_{f^{(n)}}\pi R}{2}\right) &\mbox{for $n$ odd.}
          \end{array} \right.   
          \label{fermion_mass}      
 \end{eqnarray}
Let us discuss the Yukawa interactions in this scenario as the large top quark mass plays a significant role in amplifying the quantum effects in the present study. The action of Yukawa interaction with BLTs of strength $r_y$ is written as \cite{Datta:2015aka, Datta:2016flx, Biswas:2017vhc}  
\begin{eqnarray}
\label{yukawa}
S_{Yukawa} &=& -\int d^5 x  \Big[\lambda^5_t\;\bar{\Psi}_L\widetilde{\Phi}\Psi_R 
  +r_y \;\{ \delta(y) + \delta(y-\pi R) \}\lambda^5_t\bar{\phi_L}\widetilde{\Phi}\chi_R+\textrm{h.c.}\Big].
\end{eqnarray}
The 5D coupling strength of Yukawa interaction for the third generations are represented by $\lambda^5_t$. Embedding the KK-wave-functions for fermions (given in Eqs.\;\ref{fermionexpnsn1} and \ref{fermionexpnsn2}) in the actions given in Eq.\;\ref{factn} and Eq.\;\ref{yukawa} one finds the bi-linear terms containing the doublet and singlet states of the quarks. For $n^{th}$ KK-level the mass matrix can be expressed as the following \cite{Datta:2015aka, Datta:2016flx, Biswas:2017vhc}
\begin{equation}
\label{fermion_mix}
-\begin{pmatrix}
\bar{\phi_L}^{(n)} & \bar{\phi_R}^{(n)}
\end{pmatrix}
\begin{pmatrix}
m_{f^{(n)}}\delta^{nm} & m_{t} {\mathscr{I}}^{nm}_1 \\ m_{t} {\mathscr{I}}^{mn}_2& -m_{f^{(n)}}\delta^{mn}
\end{pmatrix}
\begin{pmatrix}
\chi^{(m)}_L \\ \chi^{(m)}_R
\end{pmatrix}+{\rm h.c.}.
\end{equation}
Here, $m_t$ is identified as the mass of SM top quark while $m_{f^{(n)}}$ is obtained from the solution of the transcendental equations given in Eq.\;\ref{fermion_mass}. ${\mathscr{I}}^{nm}_1$ and ${\mathscr{I}}^{nm}_2$ are the overlap integrals which are given in the following\cite{Datta:2015aka, Datta:2016flx, Biswas:2017vhc}
  \[ {\mathscr{I}}^{nm}_1=\left(\frac{1+\frac{r_f}{\pi R}}{1+\frac{r_y}{\pi R}}\right)\times\int_0 ^{\pi R}\;dy\;
\left[ 1+ r_y \{\delta(y) + \delta(y - \pi R)\} \right] g_{R}^m f_{L}^n,\] \;\;{\rm and}\;\;\[{\mathscr{I}}^{nm}_2=\left(\frac{1+\frac{r_f}{\pi R}}{1+\frac{r_y}{\pi R}}\right)\times\int_0 ^{\pi R}\;dy\;
 g_{L}^m f_{R}^n .\]

The integral ${\mathscr{I}}^{nm}_1$ is non vanishing for both the conditions of $n=m$ and $n\neq m$ . However, for $r_y = r_f$, this integral becomes unity (when $n =m$) or zero ($n \neq m$). On the other hand, the integral ${\mathscr{I}}^{nm}_2$ is non vanishing only when  $n=m$ and becomes unity in the limit $r_y = r_f$. At this stage we would like to point out that, in our analysis we choose a condition of equality  ($r_y$=$r_f$) to elude the complicacy of mode mixing and develop a simpler form of fermion mixing matrix \cite{Jha:2014faa, Datta:2015aka, Datta:2016flx, Biswas:2017vhc}. Following this motivation, in the rest of our analysis we will maintain the equality condition\footnote{However, in general, one can choose unequal strengths of boundary terms for kinetic and Yukawa interaction for fermions.} $r_y=r_f$. 

Implying the alluded equality condition ($r_y=r_f$) the resulting mass matrix (given in Eq.\;\ref{fermion_mix}) can readily be diagonalised by following bi-unitary transformations for the left- and right-handed fields respectively \cite{Datta:2015aka, Datta:2016flx, Biswas:2017vhc}
\begin{equation}
U_{L}^{(n)}=\begin{pmatrix}
\cos\alpha_{tn} & \sin\alpha_{tn} \\ -\sin\alpha_{tn} & \cos\alpha_{tn}
\end{pmatrix},~~U_{R}^{(n)}=\begin{pmatrix}
\cos\alpha_{tn} & \sin\alpha_{tn} \\ \sin\alpha_{tn} & -\cos\alpha_{tn}
\end{pmatrix},
\end{equation}
with the mixing angle $\alpha_{tn}\left[ = \frac12\tan^{-1}\left(\frac{m_{t}}{m_{f^{(n)}}}\right)\right]$. The gauge eigen states $\Psi_L(x,y)$ and $\Psi_R(x,y)$ are related with the mass eigen states $T^1_t$ and $T^2_t$ by the given relations \cite{Datta:2015aka, Datta:2016flx, Biswas:2017vhc}

%\vspace*{-1cm}
\begin{tabular}{p{8cm}p{8cm}}
{\begin{align}
&{\phi^{(n)}_L} =  \cos\alpha_{tn}T^{1(n)}_{tL}-\sin\alpha_{tn}T^{2(n)}_{tL},\nonumber \\
&{\chi^{(n)}_L} =  \cos\alpha_{tn}T^{1(n)}_{tR}+\sin\alpha_{tn}T^{2(n)}_{tR},\nonumber
\end{align}}
%\end{eqnarray}
&
%\begin{eqnarray}
{\begin{align}
&{\phi^{(n)}_R} =  \sin\alpha_{tn}T^{1(n)}_{tL}+\cos\alpha_{tn}T^{2(n)}_{tL},\nonumber \\
&{\chi^{(n)}_R} =  \sin\alpha_{tn}T^{1(n)}_{tR}-\cos\alpha_{tn}T^{2(n)}_{tR}.
%\end{eqnarray}
\end{align}}
\end{tabular}
%\end{equation}
%\newpage
Both the physical eigen states $T^{1(n)}_t$ and $T^{2(n)}_t$ share the same mass eigen value at each KK-level. For $n^{th}$ KK-level it takes the form as $M_{t^{(n)}} \equiv \sqrt{m_{t}^{2}+m^2_{f^{(n)}}}$.

In the following we present the kinetic action (governed by $SU(2)_L \times U(1)_Y$ gauge group) of 5D gauge and scalar fields with their respective BLKTs \cite{Flacke:2008ne, Datta:2014sha, Jha:2014faa, Datta:2015aka, Datta:2016flx, Biswas:2017vhc, Dey:2016cve}
\begin{eqnarray}
S_{gauge} &=& -\frac{1}{4}\int d^5x \bigg[ W^a_{MN} W^{aMN}+r_W \left\{ \delta(y) +  \delta(y - \pi R)\right\} W^a_{\mu\nu} W^{a\mu \nu}\nonumber \\
&+& B_{MN} B^{MN}+r_B \left\{ \delta(y) +  \delta(y - \pi R)\right\} B_{\mu\nu} B^{\mu \nu}\bigg],
\label{pure-gauge}
\end{eqnarray}
\vspace*{-1cm}
\begin{eqnarray} 
S_{scalar} &=& \int d^5x \bigg[ (D_{M}\Phi)^\dagger(D^{M}\Phi) + r_\phi \left\{ \delta(y) +  \delta(y - \pi R)\right\} (D_{\mu}\Phi)^\dagger(D^{\mu}\Phi) \bigg],
\label{higgs}
\end{eqnarray}
where, $r_W$, $r_B$ and $r_\phi$ are identified as the strength of the BLKTs for the respective fields. 5D field strength tensors are written as
\begin{eqnarray}\label{ugfs}
W_{MN}^a &\equiv& (\partial_M W_N^a - \partial_N W_M^a-{\tilde{g}_2}\epsilon^{abc}W_M^bW_N^c),\\ \nonumber
B_{MN}&\equiv& (\partial_M B_N - \partial_N B_M).
\end{eqnarray}
$W^a_M (\equiv W^a_\mu, W^a_4)$ and $B_M (\equiv B_\mu, B_4)$ ($M=0,1 \ldots 4$) are represented as the 5D gauge fields corresponding to the gauge groups $SU(2)_L$ and $U(1)_Y$ respectively. 5D covariant derivative is given as $D_M\equiv\partial_M+i{\tilde{g}_2}\frac{\sigma^{a}}{2}W_M^{a}+i{\tilde{g}_1}\frac{Y}{2}B_M$, where, ${\tilde{g}_2}$ and ${\tilde{g}_1}$ represent the 5D gauge coupling constants. Here, $\frac {\sigma^{a}}{2} (a\equiv 1\ldots 3)$ and $\frac Y2$ are the generators of $SU(2)_L$ and $U(1)_Y$ gauge groups respectively. 5D Higgs doublet is represented by $\Phi=\left(\begin{array}{cc} \phi^+\\\phi^0\end{array}\right)$. Each of the gauge and scalar fields which are involved in the above actions (Eqs.\;\ref{pure-gauge} and \ref{higgs}) can be expressed in terms of appropriate KK-wave-functions as \cite{Datta:2014sha, Jha:2014faa, Datta:2015aka, Datta:2016flx, Biswas:2017vhc, Dey:2016cve}
\begin{equation}\label{Amu}
V_{\mu}(x,y)=\sum_n V_{\mu}^{(n)}(x) a^n(y),\;\;\;\;\
%\end{equation}
%\begin{equation}\label{A4}
V_{4}(x,y)=\sum_n V_{4}^{(n)}(x) b^n(y)
\end{equation}
and
\begin{equation}\label{chi}
\Phi(x,y)=\sum_n \Phi^{(n)}(x) h^n(y),
\end{equation}
where $(V_\mu, V_4)$ generically represents both the 5D $SU(2)_L$ and $U(1)_Y$ gauge bosons. 
 
Before proceeding further, we would like to make a few important remarks which could help the reader to understand the following gauge and scalar field structure as well as the corresponding KK-wave function. We know that physical neutral gauge bosons generate due to the mixing of $B$ and $W^3$ fields and hence the KK-decomposition of neutral gauge bosons become very intricate in the present extra dimensional scenario because of the existence of two types of mixings both at the bulk as well as on the boundary. Therefore, in this situation without the condition $r_W=r_B$, it would be very difficult to diagonalise the bulk and boundary actions simultaneously by the same 5D field redefinition\footnote{However, in general one can proceed with $r_W\neq r_B$, but in this situation the mixing between $B$ and $W^3$ in the bulk and on the boundary points produce off-diagonal terms in the neutral gauge boson mass matrix.}. Hence, in the following we will sustain the equality condition $r_W=r_B$ \cite{Datta:2014sha, Jha:2014faa, Datta:2015aka, Datta:2016flx, Biswas:2017vhc, Dey:2016cve}. Consequently, similar to the mUED scenario, one obtains the same structure of mixing between KK-excitations of the neutral component of the gauge fields (i.e., the mixing between $W^{3(n)}$ and $B^{(n)}$) in nmUED scenario. Therefore, the mixing between $W^{3(1)}$ and $B^{(1)}$ (i.e., the mixing at the first KK-level) gives the $Z^{(1)}$ and $\gamma^{(1)}$. This $\gamma^{(1)}$ (first excited KK-state of photon) is absolutely stable by the conservation of KK-parity and it possesses the lowest mass among the first excited KK-states in the nmUED particle spectrum. Moreover, it can not decay to pair of SM particles. Therefore, this $\gamma^{(1)}$ has been played the role of a viable DM candidate in this scenario \cite{Datta:2013nua}.  

In the following, we have given the gauge fixing action (contains a generic BLKT parameter $r_V$ for gauge bosons) appropriate for nmUED model\cite{Datta:2014sha, Jha:2014faa, Datta:2015aka, Datta:2016flx, Biswas:2017vhc, Dey:2016cve} 
\begin{eqnarray} 
S_{gauge\;fixing} &=& -\frac{1}{\xi _y}\int d^5x\Big\vert\partial_{\mu}W^{\mu +}+\xi_{y}(\partial_{y}W^{4+}+iM_{W}\phi^{+}\{1 + r_{V}\left( \delta(y) + \delta(y - \pi R)\right)\})\Big \vert ^2 \nonumber \\&&-\frac{1}{2\xi_y}\int d^5x [\partial_{\mu}Z^{\mu}+\xi_y(\partial_{y}Z^{4}-M_{Z}\chi\{1+ r_V( \delta(y) +  \delta(y - \pi R))\})]^2\nonumber \\
&-&\frac{1}{2\xi_y}\int d^5x [\partial_{\mu}A^{\mu}+\xi_y\partial_{y}A^{4}]^2, 
\label{gauge-fix}
\end{eqnarray}
where $M_W(M_Z)$ is the mass of SM $W^\pm (Z)$ boson.
For a detailed study on gauge fixing action/mechanism in nmUED we refer to \cite{Datta:2014sha}. The above action (given in Eq.\;\ref{gauge-fix}) is somewhat intricate and at the same time very crucial for this nmUED scenario where we will calculate one loop diagrams (required for present calculation) in Feynman gauge. In the presence of the BLKTs the Lagrangian leads to a non-homogeneous weight function for the fields with respect to the extra dimension. This inhomogeneity compels us to define a $y$-dependent gauge fixing parameter $\xi_y$ as \cite{Datta:2014sha, Jha:2014faa, Datta:2015aka, Datta:2016flx, Biswas:2017vhc, Dey:2016cve}
\begin{equation}\label{gf_para}
\xi =\xi_y\,(1+ r_V\{ \delta(y) +  \delta(y - \pi R)\}),
\end{equation}
where $\xi$ is not dependent on $y$. This relation can be treated as {\em renormalisation} of the gauge fixing parameter since the BLKTs are in some sense played the role of counterterms taking into account the unknown ultraviolet contribution in loop calculations. In this sense, 
$\xi_y$ is the bare gauge fixing parameter while $\xi$ can be seen as the renormalized gauge fixing parameter taking the values $0$ (Landau gauge),
$1$ (Feynman gauge) or $\infty$ (Unitary gauge) \cite{Datta:2014sha}.

In the present scenario appropriate gauge fixing procedure enforces the condition $r_V=r_\phi$~\cite{Datta:2014sha, Jha:2014faa, Datta:2015aka, Datta:2016flx, Biswas:2017vhc, Dey:2016cve}. Consequently, KK-masses for the gauge and the scalar field are equal ($m_{V^{(n)}}(=m_{\phi^{(n)}})$) and satisfy the same transcendental equation (Eq.~\ref{fermion_mass}). At the $n^{th}$ KK-level the physical gauge fields ($W^{\mu (n)\pm}$) and charged Higgs ($H^{(n)\pm}$) share the same\footnote{Similarly one can find the mass eigen values for the KK-excited $Z$ boson and pseudo scalar $A$. Moreover, their mass eigen values are also identical to each other at any KK-level. For example at $n^{th}$ KK-level it takes the form as $\sqrt{M_{Z}^{2}+m^2_{V^{(n)}}}$. } mass eigen value and is given by\cite{Datta:2014sha, Jha:2014faa, Datta:2015aka, Datta:2016flx, Biswas:2017vhc, Dey:2016cve} 
\be\label{MWn}
M_{W^{(n)}} = \sqrt{M_{W}^{2}+m^2_{V^{(n)}}}\;.
\ee 
Moreover, in the ’t-Hooft Feynman gauge, the mass of Goldstone bosons ($G^{(n)\pm}$) corresponding to the gauge fields $W^{\mu (n)\pm}$ has the same value $M_{W^{(n)}}$\cite{Datta:2014sha, Jha:2014faa, Datta:2015aka, Datta:2016flx, Biswas:2017vhc, Dey:2016cve}.

Additionally, we would like to mention that, as in the present article we are dealing with a process that involves off-shell amplitude, hence we need to use the method of background fields \cite{Deshpande:1982mi, Buras:2003mk}. We have already mentioned that the same decay process has already been calculated in the article \cite{Buras:2003mk} in the context of 5D UED and further the authors have also used the same background fields. For this reason, in the {\it Appendix A} of that article \cite{Buras:2003mk} the authors have discussed the background field method and also given the corresponding prescription for the 5D UED scenario. We can readily adopt this prescription in the present nmUED scenario because the basic structure of both these models are similar. We hence refrain from providing the details of this method in the present scenario. However, using that prescription (given in \cite{Buras:2003mk}) we can easily evaluate the Feynman rules necessary for our present calculation. In Appendix \ref{fyerul} we give the necessary Feynman rules derived for the 5D background field method in the 5D nmUED scenario in Feynman gauge.
 
Up to this we provide the relevant information of the present scenario. At this stage it is important to mention that the interactions for our calculation can be evaluated by integrating out the 5D action over the extra space-like dimension ($y$) after plugging the appropriate $y$-dependent KK-wave-function for the respective fields in 5D action. As a consequence some of the interactions are modified by so called overlap integrals with respect to their mUED counterparts. The expressions of the overlap integrals have been given in Appendix \ref{fyerul}. For further information of these overlap integrals we refer the reader to check \cite{Datta:2015aka}.

\section{\boldmath$B\to X_{\lowercase{s}} \lowercase{\ell}^+\lowercase{\ell}^-$ in nmUED}
         \label{sec:Heff:BXsee:nlo}
The semileptonic inclusive decay $B\to X_s \ell^+\ell^-$ is quite suppressed in the SM, however it is very compelling for finding NP signature. Therefore, several $B$-physics experimental collaborations (Belle, Babar) have been involved to measure several observables (mainly decay branching ratio, Forward-Backward asymmetry) associated with this decay process. In the context of SM, the dominant perturbative
contribution has been evaluated in \cite{Grinstein:1988me} and later two loop QCD corrections\footnote{Research regarding higher order perturbative contributions has
been studied extensively and has already reached at a high level accuracy. For example, one can find NNLO QCD corrections in \cite{Bobeth:1999mk} and including QED 
corrections in \cite{Bobeth:2003at, Huber:2005ig}. Moreover, updated analysis
of all angular observables in the $B\to X_s \ell^+\ell^-$ decay has been given in \cite{Huber:2015sra}. It also contains all available perturbative NNLO QCD, NLO QED corrections and includes subleading power corrections.} have been
described in the refs. \cite{Misiak:1992bc, Buras:1994dj}. Since in this particular decay mode, a lepton-antilepton pair is present, therefore, more structures contribute to the decay rate and some subtleties arise in theoretical description for this process. For the decay to be dominated by perturbative contributions then one has to eliminate $c\bar{c}$ resonances that show up as large peaks in the di-lepton invariant mass spectrum by judicious choice of kinematic cuts. Consequently this leads to ``perturbative di-lepton invariant mass windows'', namely the low di-lepton mass region $1~{\rm GeV^2} < q^2 < 6~{\rm GeV^2}$, and also the high di-lepton mass region with $q^2 > 14.4~{\rm GeV^2}$. 

In this section we will describe the details of the calculation of the branching
ratio and Forward-Backward asymmetry of $B\to X_s\ell^+\ell^-$ in nmUED model. Since the basic gauge structure of the present nmUED model is similar to SM, therefore, leading order (LO)
contributions to electroweak dipole operators are one loop suppressed as in SM. However, in the present model due to the presence of large number of KK-particles we encounter more one loop diagrams in comparison to SM. Hence, we will 
evaluate the total contributions of these KK-particles to the electroweak dipole operators and just simply
add them to SM contribution. With this spirit following the same technique of the ref.\cite{Buras:2003mk} we
will evaluate the relevant WCs of the electroweak dipole operators at the LO level. Then following the prescription
of as given in \cite{Misiak:1992bc, Buras:1994dj} we will include NLO QCD correction to the concerned decay process.
%\vspace*{-0.3cm}
\subsection{Effective Hamiltonian for \boldmath$B\to X_s\ell^+\ell^-$} 
\begin{figure}[th!]
\begin{center}
\includegraphics[scale=0.80,angle=0]{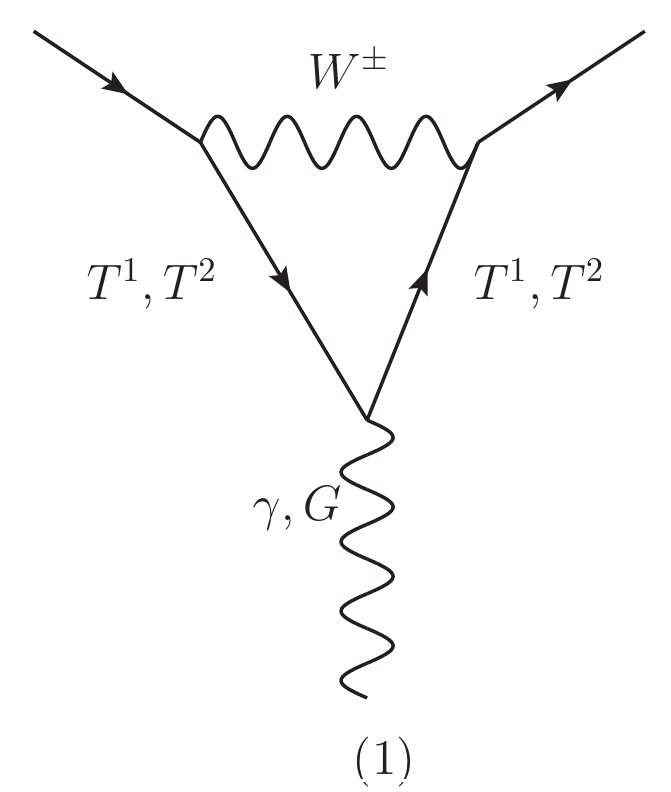}
\includegraphics[scale=0.80,angle=0]{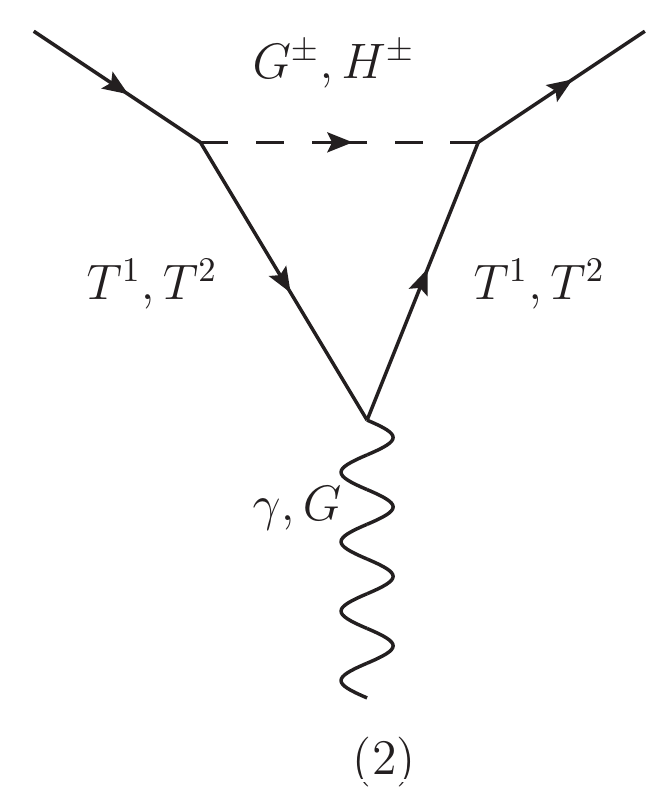}
\includegraphics[scale=0.80,angle=0]{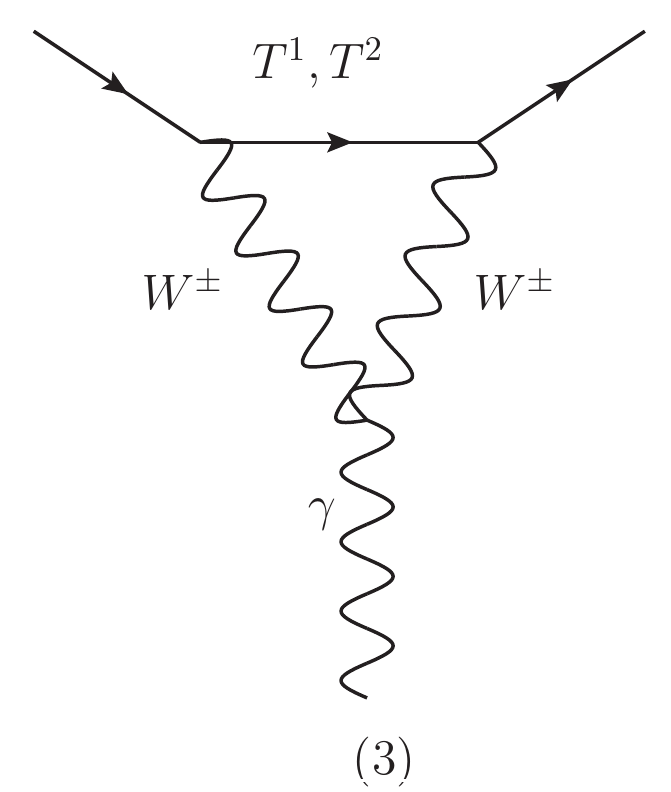}\\
\includegraphics[scale=0.80,angle=0]{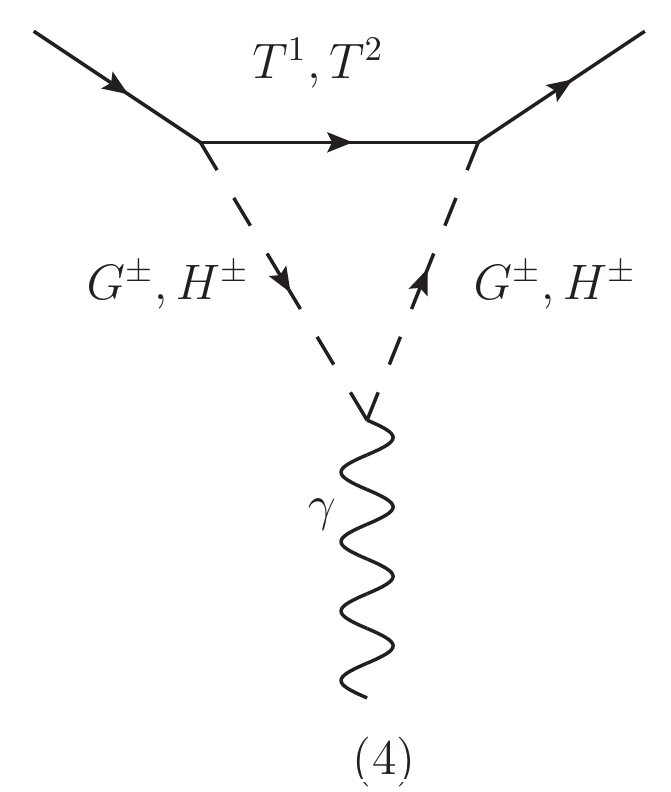}
\includegraphics[scale=0.80,angle=0]{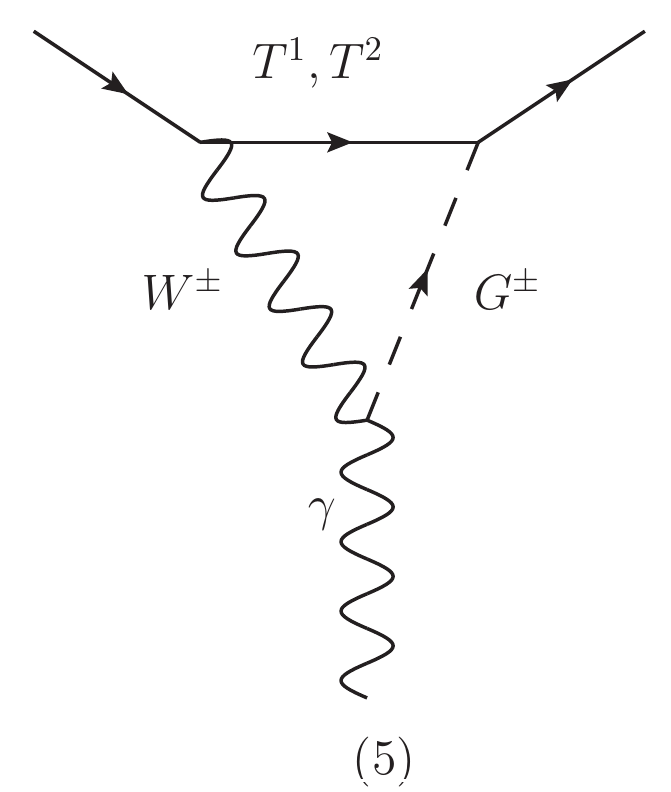}
\includegraphics[scale=0.80,angle=0]{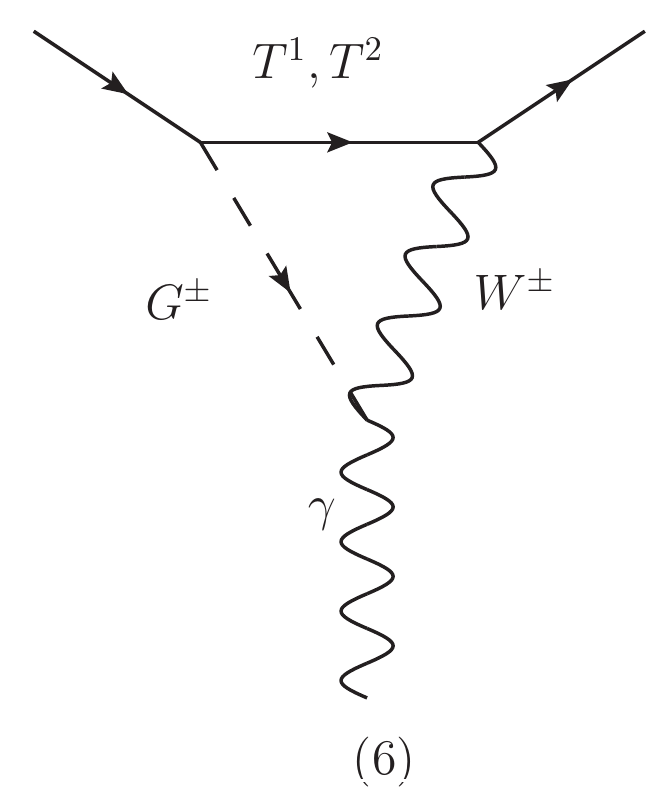}
\caption{Relevant electroweak penguin diagrams contributing to the decay of $B\to X_s\ell^+\ell^-$.}
\label{magnetic_pen}
\end{center}
\end{figure}

The effective Hamiltonian for the decay $B\to X_s\ell^+\ell^-$ at hadronic scales $\mu=\mathcal{O}(m_b)$ can be written as \cite{Buras:2003mk}
\begin{equation} \label{Heff2_at_mu}
{\cal H}_{\rm eff}(b\to s \ell^+\ell^-) =
{\cal H}_{\rm eff}(b\to s\gamma)  - \frac{G_{\rm F}}{\sqrt{2}} V_{ts}^* V_{tb}
\left[ C_{9V}(\mu) Q_{9V}+
C_{10A}(M_W) Q_{10A}    \right]\,,
\end{equation}
where $G_F$ represents the Fermi constant and $V_{ij}$ are the elements of Cabibbo-Kobayashi-Maskawa (CKM) matrix. In the above expression (Eq.\;\ref{Heff2_at_mu}) apart from the relevant operators\footnote{The explicit form of the effective Hamiltonian for $b\to s \gamma$ is given in \cite{Buras:2003mk, Datta:2016flx}.} for $B\to X_s\gamma$ there are two new operators \cite{Buras:2003mk}
\begin{equation}\label{Q9V}
Q_{9V}    = (\bar{s} b)_{V-A}  (\bar{\ell}\ell)_V\,,         
\qquad
Q_{10A}  =  (\bar{s} b)_{V-A}  (\bar{\ell}\ell)_A\,,
\end{equation}
where $V$ and $A$ refer to vector and axial-vector current respectively. They are produced via the electroweak penguin diagrams shown in Fig.~\ref{magnetic_pen} and the other relevant Feynman diagrams needed to maintain gauge invariance (for nmUED scenario) has been given in \cite{Datta:2015aka}.

For the purpose of convenience the above WCs (given in Eq.\;\ref{Heff2_at_mu}) can be defined in terms of two new coefficients $\tilde C_{9}$ and $\tilde C_{10}$  as \cite{Buras:2003mk, Buras:1994dj}
\begin{eqnarray} \label{C9_10}
C_{9V}(\mu) &=& \frac{\alpha}{2\pi} \tilde C_9(\mu), \\
C_{10A}(\mu) &=&  \frac{\alpha}{2\pi} \tilde C_{10}(\mu), 
\end{eqnarray}
where,
\begin{equation}
\tilde C_{10}(\mu) = - \frac{Y(x_t, r_f, r_V, R^{-1})}{\sin^2\theta_{w}}\;.
\end{equation}
The function $Y(x_t, r_f, r_V, R^{-1})$ in the context of nmUED scenario has been calculated in \cite{Datta:2015aka}. $\theta_w$ is the Weinberg angle and $\alpha$ represents the fine structure constant. The operator $Q_{10A}$ does not evolve under QCD renormalisation and  its coefficient is independent of $\mu$. On the other hand using the results of NLO QCD corrections to $\tilde C_{9}(\mu)$ in the SM given in \cite{Misiak:1992bc, Buras:1994dj} we can readily obtain this coefficient in the present nmUED model under the naive dimensional regularisation (NDR) renormalisation scheme as
\begin{eqnarray}\label{c9_eff}
\Ctilde_9^{\rm eff}(q^2)&=&\Ctilde_9^{\rm NDR}\tilde{\eta}\left(\frac{q^2}{m^2_b}\right)+h\left(z,\frac{q^2}{m^2_b}\right)\left(3C^{(0)}_1+C^{(0)}_2+3C^{(0)}_3+C^{(0)}_4+3C^{(0)}_5+C^{(0)}_6\right) \\ \nonumber 
&&-\frac 12 h\left(1,\frac{q^2}{m^2_b}\right)\left(4C^{(0)}_3+4C^{(0)}_4+3C^{(0)}_5+C^{(0)}_6\right)-\frac 12 h\left(0,\frac{q^2}{m^2_b}\right)\left(C^{(0)}_3+4C^{(0)}_4\right)  \\ \nonumber 
&&+\frac 29\left(3C^{(0)}_3+C^{(0)}_4+3C^{(0)}_5+C^{(0)}_6\right),
\end{eqnarray}
where,
\begin{equation}\label{C9tilde}
\Ctilde_9^{\rm NDR}(\mu)  =  
P_0^{\rm NDR} + \frac{Y(x_t, r_f, r_V, R^{-1})}{\sin^2\theta_{w}} -4 Z(x_t, r_f, r_V, R^{-1}) + P_E E(x_t, r_f, r_V, R^{-1})\;.
\end{equation}
The value\footnote{The analytic formula for $P_0^{\rm NDR}$ has been given in
\cite{Buras:1994dj}.} of $P_0^{\rm NDR}$ is $2.60\pm 0.25$ \cite{Buras:2003mk} and $P_E$ is ${\cal O}(10^{-2})$ \cite{Buras:1994dj}. Using the relation  given in \cite{Buras:1994dj, Buras:2003mk} we can express the function $Z$ in nmUED scenario as 
\begin{equation}
\label{Zfunction} 
Z(x_t, r_f, r_V, R^{-1})=C(x_t, r_f, r_V, R^{-1})+\frac 14 D(x_t, r_f, r_V, R^{-1})\;,
\end{equation}
while the function $C(x_t, r_f, r_V, R^{-1})$ for nmUED scenario has been calculated in \cite{Datta:2015aka}. The function $\tilde{\eta}$ given in the Eq.\;\ref{c9_eff} represents single gluon corrections to the matrix element $Q_9$ and it takes the form as \cite{Buras:1994dj} 
\begin{eqnarray}
\tilde{\eta}\left(\frac{q^2}{m^2_b}\right)=1+\frac{\alpha_s}{\pi}\omega\left(\frac{q^2}{m^2_b}\right),
\end{eqnarray}
where $\alpha_s$ is the QCD fine structure constant. The explicit form of the functions $\omega$, $h$ and other WCs (e.g., given in Eq.\;\ref{c9_eff}) required for the present decay process have been given in Appendix \ref{NDR}.  The functions $D(x_t, r_f, r_V, R^{-1})$  and $E(x_t, r_f, r_V, R^{-1})$  which we evaluate in this article are  given in the following
\begin{equation}
D(x_t, r_f, r_V, R^{-1})=D_0(x_t)+
\sum_{n=1}^\infty D_n(x_t,x_{f^{(n)}},x_{V^{(n)}})\;,
\label{dsum}
\end{equation}
and 
\begin{equation}
E(x_t, r_f, r_V, R^{-1})=E_0(x_t)+
\sum_{n=1}^\infty E_n(x_t,x_{f^{(n)}},x_{V^{(n)}})\;,
\label{esum}
\end{equation}
with $x_t=\frac{m^2_t}{M^2_W}$, $x_{V^{(n)}}=\frac{m^2_{V^{(n)}}}{M^2_W}$ and $x_{f^{(n)}}=\frac{m^2_{f^{(n)}}}{M^2_W}$. $m_{V^{(n)}}$ and $m_{f^{(n)}}$ can be obtained from transcendental equation given in Eq.\;\ref{fermion_mass}. The functions $D_0(x_t)$ and $E_0(x_t)$ are the corresponding SM contributions at the electroweak scale \cite{Buras:2003mk, Misiak:1992bc, Buras:1994dj, Inami:1980fz, Buras:1994qa}
\begin{equation}\label{DSM}
D_0(x_t)=-{4\over9}\ln x_t+{{-19x_t^3+25x_t^2}\over{36(x_t-1)^3}}
+{{x_t^2(5x_t^2-2x_t-6)} \over{18(x_t-1)^4}}\ln x_t~,
\end{equation}
\begin{equation}\label{ESM}
E_0(x_t)=-{2\over 3}\ln x_t+{{x_t^2(15-16x_t+4x_t^2)}\over{6(1-x_t)^4}}
\ln x_t+{{x_t(18-11x_t-x_t^2)} \over{12(1-x_t)^3}}~.
\end{equation}

Now we will depict the nmUED contribution to the electroweak penguin diagrams. We have already mentioned that the KK-masses and  couplings involving KK-excitations are non-trivially modified with respect to their UED counterparts due to the presence of different BLTs in the nmUED action. Therefore, it would not be possible to obtain the expressions of $D$ and $E$ functions in nmUED simply by rescaling the results of UED model \cite{Buras:2003mk}. Consequently, we have evaluated the functions $D_n(x_t,x_{f^{(n)}},x_{V^{(n)}})$ and $E_n(x_t,x_{f^{(n)}},x_{V^{(n)}})$ {\it independently} for the nmUED scenario.
These functions ($D_n$ and $E_n$) represent the KK-contributions for $n^{th}$ KK-mode which are computed from the electroweak penguin diagrams (given in Fig.\;\ref{magnetic_pen}) in nmUED model for photon and gluon respectively. Furthermore, it is quite evident from Eqs.\;\ref{dn} 
 and \ref{en} that they are remarkably different from that of the UED expression (given in Eqs.\;3.31 and 3.32 of ref.\;\cite{Buras:2003mk}). However, from our expressions (given in Eqs.\;\ref{dn} and \ref{en}) we can reconstruct the results of UED version (given in the Eqs.\;3.31 and 3.32 of the ref.\;\cite{Buras:2003mk}) if we set the boundary terms to zero i.e., $r_f, r_V = 0$.

%\vspace*{-0.5cm}

To this end, we would like to mention that in our calculation of one loop penguin diagrams (in order to measure the contributions of KK-excitation to the decay of $B\rightarrow X_s\ell^+\ell^-$) we consider only those interactions which couple a zero-mode field to a pair of KK-excitations carrying equal KK-number. Although, in nmUED scenario due to the KK-parity conservation one can also have non-zero interactions involving KK-excitations with  KK-numbers $n, m~{\rm and}~p$ where $n+m+p$ is an even integer. However, we have explicitly checked that the final results would not change remarkably even if one considers the contributions of all the possible off-diagonal interactions \cite{Jha:2014faa, Datta:2015aka, Datta:2016flx}.

%\newpage
For the $n^{th}$ KK-level the electroweak {\it photon} penguin function (which is obtained from  penguin diagrams given in Fig.\;\ref{magnetic_pen}) takes the form as
%\vspace*{-0.2cm}
\begin{eqnarray}\label{dn}
D_n(x_t,x_{f^{(n)}},x_{V^{(n)}})&=&\frac 23E_n(x_t,x_{f^{(n)}},x_{V^{(n)}})-\frac{1}{36(-1+x_{f^{(n)}}-x_{V^{(n)}})^4}\\ \nonumber
&&\bigg[(-1+x_{f^{(n)}}-x_{V^{(n)}})\bigg\{-2(I^n_1)^2\bigg(43x^2_{f^{(n)}}-65x_{f^{(n)}}(1+x_{V^{(n)}})\\ \nonumber
&&+16(1+x_{V^{(n)}})^2\bigg)+(I^n_2)^2\bigg(11x^2_{f^{(n)}}-7x_{f^{(n)}}(1+x_{V^{(n)}})\\ \nonumber
&&+2(1+x_{V^{(n)}})^2
\bigg)\bigg\}-6x^2_{f^{(n)}}\bigg\{(I^n_2)^2x_{f^{(n)}}+2(I^n_1)^2\\ \nonumber
&&\bigg(6-5x_{f^{(n)}}+6x_{V^{(n)}}\bigg)\bigg\}\ln\bigg(\frac{x_{f^{(n)}}}{1+x_{V^{(n)}}}\bigg) 
\bigg]\\ \nonumber
&&+\frac{1}{36(-1+x_t+x_{f^{(n)}}-x_{V^{(n)}})^4}\bigg[(-1+x_t+x_{f^{(n)}}-x_{V^{(n)}})\bigg\{\\ \nonumber
&&(I^n_1)^2\Bigg(11x^3_t+x^2_{f^{(n)}}(-86+11x_t)-
x^2_t(93+7x_{V^{(n)}})+32(1+x_{V^{(n)}})^2\\ \nonumber
&&+2x_t(1+x_{V^{(n)}})(66+x_{V^{(n)}})+x_{f^{(n)}}\bigg(x_t(-179+22x_t-7x_{V^{(n)}})\\ \nonumber
&&130(1+x_{V^{(n)}})\bigg)\Bigg)+(I^n_2)^2\Bigg(11x^2_{f^{(n)}}+11x^2_t-7x_t(1+x_{V^{(n)}})\\ \nonumber
&&+2(1+x_{V^{(n)}})^2+x_{f^{(n)}}\bigg(22x_t-7(1+x_{V^{(n)}})\bigg)
\Bigg)\bigg\}-6(x_t+x_{f^{(n)}})^2\\ \nonumber
&&\bigg\{(I^n_2)^2(x_t+x_{f^{(n)}})
+(I^n_1)^2\bigg((x_t+x_{f^{(n)}})(-10+x_t)\\ \nonumber
&&+12(1+x_{V^{(n)}})\bigg)\bigg\}\ln\bigg(\frac{x_t+x_{f^{(n)}}}{1+x_{V^{(n)}}}\bigg) 
\bigg], 
\end{eqnarray}
while the function $E_n(x_t,x_{f^{(n)}},x_{V^{(n)}})$ is regarded as the corresponding contribution for $gluon$ penguins given by the first two diagrams of Fig.\;\ref{magnetic_pen}. The expression of the function $E_n(x_t,x_{f^{(n)}},x_{V^{(n)}})$ in nmUED is given as the following

%\vspace*{-0.2cm}
\begin{eqnarray}\label{en}
E_n(x_t,x_{f^{(n)}},x_{V^{(n)}})&=&-\frac{1}{36(-1+x_{f^{(n)}}-x_{V^{(n)}})^4}\bigg[(-1+x_{f^{(n)}}-x_{V^{(n)}})\bigg\{\\ \nonumber
&&(I^n_1)^2\bigg(50x^2_{f^{(n)}}-58x_{f^{(n)}}(1+x_{V^{(n)}})-4(1+x_{V^{(n)}})^2\bigg)+(I^n_2)^2\\ \nonumber
&&\bigg(7x^2_{f^{(n)}}-29x_{f^{(n)}}(1+x_{V^{(n)}})+16(1+x_{V^{(n)}})^2
\bigg)\bigg\}-6(1+x_{V^{(n)}})\\ \nonumber
&&\bigg\{(I^n_2)^2(1+x_{V^{(n)}})(2-3x_{f^{(n)}}+2x_{V^{(n)}})
+2(I^n_1)^2\\ \nonumber
&&\bigg(6x^2_{f^{(n)}}-9x_{f^{(n)}}(1+x_{V^{(n)}})+2(1+x_{V^{(n)}})^2\bigg)\bigg\}\ln\bigg(\frac{x_{f^{(n)}}}{1+x_{V^{(n)}}}\bigg) 
\bigg]\\ \nonumber
&&+\frac{1}{36(-1+x_t+x_{f^{(n)}}-x_{V^{(n)}})^4}\bigg[(-1+x_t+x_{f^{(n)}}-x_{V^{(n)}})\bigg\{\\ \nonumber
&&(I^n_1)^2\Bigg(7x^3_t+x^2_{f^{(n)}}(50+7x_t)+
x^2_t(21-29x_{V^{(n)}})-4(1+x_{V^{(n)}})^2\\ \nonumber
&&+2x_t(1+x_{V^{(n)}})(-21+8x_{V^{(n)}})+x_{f^{(n)}}\bigg(-58+71x_t+14x^2_t\\ \nonumber
&&-29(2+x_t)x_{V^{(n)}}\bigg)\Bigg)+(I^n_2)^2\Bigg(7x^2_{f^{(n)}}+7x^2_t-29x_t(1+x_{V^{(n)}})\\ \nonumber
&&+16(1+x_{V^{(n)}})^2+x_{f^{(n)}}\bigg(14x_t-29(1+x_{V^{(n)}})\bigg)
\Bigg)\bigg\}-6(1+x_{V^{(n)}})\\ \nonumber
&&\bigg\{(I^n_2)^2(1+x_{V^{(n)}})(2-3x_{f^{(n)}}-3x_t+2x_{V^{(n)}})
+(I^n_1)^2\Bigg(12x^2_{f^{(n)}}\\ \nonumber
&&-3x^2_t(-3+x_{V^{(n)}})+2x_t(-8+x_{V^{(n)}})(1+x_{V^{(n)}})+4(1+x_{V^{(n)}})^2\\ \nonumber
&&-3x_{f^{(n)}}\bigg(6-7x_t+(6+x_t)x_{V^{(n)}}\bigg)\Bigg)\bigg\}\ln\bigg(\frac{x_t+x_{f^{(n)}}}{1+x_{V^{(n)}}}\bigg) 
\bigg].
\end{eqnarray}
In the above expressions, $I^n_1$ and $I^n_2$ represent overlap integrals whose analytic forms have been given in the Appendix \ref{fyerul} (see Eqs.\;\ref{i1} and \ref{i2}).
\vspace*{-0.5cm}
\subsection{The Differential Decay Rate}
\label{sec:Heff:BXsee:nlo:rate}
We are now in a stage where on the basis of effective Hamiltonian given in Eq.\;\ref{Heff2_at_mu} we can readily define the differential decay rate in the NDR scheme \cite{Misiak:1992bc,Buras:1994dj}
\be \label{rateee}
R(q^2) \equiv \frac{1}{\Gamma(b \to c e\bar\nu)}\frac{{d}\Gamma (b \to s \ell^+\ell^-)}{d q^2} \, 
 = \frac{\alpha^2}{4\pi^2}
\left|\frac{V^*_{ts}V_{tb}}{V_{cb}}\right|^2 \frac{\left(1-\frac{q^2}{m^2_b}\right)^2}{f(z)\kappa(z)}
U(q^2).
\ee
Here, 
\be \label{fz}
f(z)=1-8z^2+8z^6-z^8-24z^4\ln(z),
\ee
is the phase-space factor and
\be \label{kz}
\kappa(z)\simeq 1-\frac{2\alpha_s(\mu)}{3\pi}\bigg[\bigg(\pi^2-\frac{31}{4}\bigg)(1-z)^2+\frac 32\bigg],
\ee
represents the single gluon QCD correction to $b\to c e\bar\nu$ decay \cite{Cabibbo:1978sw,Kim:1989ac} with $z=\frac{m_c}{m_b}$. The function $U(q^2)$  is expressed as
%$\kappa(z)=0.88$
%In the above Eq.\;\ref{rateee}
\be\label{US} 
U(q^2)=
\left(1+\frac{2q^2}{m^2_b}\right)\left(|\Ctilde_9^{\rm eff}(q^2)|^2 + |\Ctilde_{10}|^2\right) + 
4 \left( 1 + \frac{2m^2_b}{q^2}\right) |C_{7\gamma}^{(0){\rm eff}}|^2 + 12
C_{7\gamma}^{(0){\rm eff}} \ \RE\,\Ctilde_9^{\rm eff}(q^2),
\ee
where, $\Ctilde_9^{\rm eff}(q^2)$ is given in Eq.\;\ref{c9_eff}. The explicit formula for $C_{7\gamma}^{(0){\rm eff}}$ is shown in the Appendix \ref{NDR}. Among the several terms given in Eq.\;\ref{US},\;\ $|\Ctilde_9^{\rm eff}(q^2)|^2$ is 
almost similar to that of the SM, $|\Ctilde_{10}|^2$ is appreciably
enhanced, however the last two terms are suppressed. 
Furthermore, the last term in Eq.\;\ref{US} is negative and hence its
suppression results are responsible for an enhancement of $U(q^2)$ in addition to the 
one due to $\Ctilde_{10}$. Using Eq.\;\ref{rateee}, one can easily evaluate branching ratio for the present decay process for a given range of $q^2$. In the numerical calculations we will use the value 0.104 for ${\rm Br}(B\to X_c e\bar\nu)_{\rm exp}$.

\subsection{Forward-Backward Asymmetry}
\label{FBA}
For the present decay process $B\to X_s\ell^+\ell^-$ another observable called Forward-Backward asymmetry could be instrumental for the detection of NP scenario. It is non-zero only at the NLO level. The unnormalised expression is given as
\cite{Ali:1991is}
\begin{eqnarray}\label{unABF}
\bar{A}_{FB}(q^2)&\equiv&\frac{1}{\Gamma
(b \to c e\bar\nu)} \int_{-1}^1 d \cos \theta_\ell \frac{ d^2
  \Gamma (b \to s \ell^+\ell^-)} { dq^2 d\cos \theta_\ell} {\rm sgn}
(\cos \theta_l), \\ 
&=& -3\frac{\alpha^2}{4\pi^2}
\left|\frac{V^*_{ts}V_{tb}}{V_{cb}}\right|^2 \frac{\left(1-\frac{q^2}{m^2_b}\right)^2}{f(z)\kappa(z)} \tilde C_{10}
\left[\frac{q^2}{m^2_b} \RE\,\Ctilde_9^{\rm eff}(q^2)
+2 C_{7\gamma}^{(0){\rm eff}}\right].
\end{eqnarray}
Here, $\theta_\ell$ represents the angle
of the $\ell^+$ with respect to $b$-quark direction in the centre-of-mass system of the
di-lepton pair. The normalised form can be expressed as
\begin{eqnarray}
\label{nAFB}
A_{FB}=\frac{\bar{A}_{FB}(q^2)}{R(q^2)},
\end{eqnarray}
while the global Forward-Backward asymmetry in a region $q^2\in[a,\;b]\;{\rm GeV}^2$ can be defined as \cite{Feng:2016wph, Lunghi:1999uk}
\begin{eqnarray}
%%%%%%%%%%%%%%%%%%%%%%%%%%%%%%%%%%%%%%%%%%%%%%%%%%%
&&A_{_{FB}}\Big|_{q^2\in[a,\;b]\;{\rm GeV}^2}
={\int_a^b dq^2\bar{A}_{_{FB}}(q^2)\over\int_a^bdq^2R(q^2)}\;.
%%%%%%%%%%%%%%%%%%%%%%%%%%%%%%%%%%%%%%%%%%%%%%%%%%%
\label{global-AFB}
\end{eqnarray}
In the following section we will present the numerical estimation of these observables for the allowed parameter space in nmUED scenario.
 
\section{Analysis and results}\label{anls}
The effective Hamiltonian (given in Eq.\;\ref{Heff2_at_mu}) required for the decay $B\rightarrow X_s\ell^+\ell^-$ contains different WCs and in our analysis we evaluate KK-contributions to each of these coefficients at each KK-level. In this article, for the first time we have calculated the KK-contributions to the coefficients of electroweak dipole operators in the nmUED scenario. The functions $D_n(x_t,x_{f^{(n)}},x_{V^{(n)}})$ (given in Eq.~\ref{dn}) and $E_n(x_t,x_{f^{(n)}},x_{V^{(n)}})$ (given in Eq.~\ref{en}) represent the $n^{th}$ level KK-contributions to the coefficients for the dipole operators for photon and gluon respectively. These functions ($D_n$ and $E_n$) depend on gauge boson as well as fermion KK-masses\footnote{We use $M_W=80.38$ GeV for SM $W^\pm$ gauge boson mass and $m_t=173.1$ GeV for SM top quark mass as given in ref.\;\cite{Tanabashi:2018oca}.} in the nmUED scenario. Furthermore, other coefficients needed for the concerned decay process in nmUED scenario have been given in our previous articles \cite{Datta:2015aka, Datta:2016flx}. At this point we would like to mention that, considering the analysis of the effect of SM Higgs mass on vacuum stability in UED model \cite{Datta:2012db}, we sum the KK-contributions up to 5 KK-levels\footnote{Analysis in earlier articles used 20-30 KK-levels while adding up the contributions from KK-modes.} and finally we add up the total KK-contributions with the SM counterpart. In fact, we have explicitly checked that the numerical values would not differ remarkably as the sum over the KK-modes, in this case, is converging\footnote{The summation of KK-contribution is convergent in UED type models with one extra space-like dimension, as far as one loop calculation is concerned\cite{Dey:2004gb}.} in nature. More specifically, during the calculation of loop diagrams, the summation of KK-levels becomes saturated after consideration of a certain number of KK-levels. Consequently, the final results would not change significantly whether we consider 5 KK-levels or 20 KK-levels during the evaluation of KK-contributions for the loop diagrams. In support of our assumption, at the end of the following subsection, we will present two tables (Tables \ref{sum-kk_low} and \ref{sum-kk_high}) which will ensure the insensitivity on the number of KK-levels in summation. 

\subsection{Constraints and choice of range of BLT parameters}
Here we briefly discuss the following constraints that have been imposed in our analysis.
\begin{itemize}
\item Several rare decay processes, for example $B_s\rightarrow \mu^+\mu^-$ and $B\rightarrow X_s\gamma$ have always been very crucial for searching any favourable kind of NP scenario. The latest experimental values for branching ratios of these processes are given in the following
\begin{table}[H]
\begin{center}
\begin{tabular}{|c|c|}
\hline
Process & Experimental value of branching ratio\\ \hline
$B_s\rightarrow \mu^+\mu^-$ & $(2.8^{+0.8}_{-0.7})\times 10^{-9}$ \cite{Aaboud:2018mst}\\ \hline
$B\rightarrow X_s\gamma$ & $(3.32\pm 0.16)\times 10^{-4}$ \cite{Amhis:2016xyh}\\ \hline
\end{tabular}
\caption{Experimental values for branching ratios of $B_{s} \rightarrow \mu^+ \mu^-$ and $B\rightarrow X_s\gamma$.}
\label{t:4}
\end{center}
\end{table}
In the context of nmUED scenario, thorough analyses on the above mentioned rare decay processes have been performed in refs.~\cite{Datta:2015aka} and \cite{Datta:2016flx} respectively. Using the expressions of ${\rm Br}(B_s\rightarrow \mu^+\mu^-)$ and ${\rm Br}(B\rightarrow X_s\gamma)$ given in \cite{Datta:2015aka} and \cite{Datta:2016flx} we have treated the branching ratios of these rare decay processes as constraints in our present analysis.

\item Electroweak precision test (EWPT) is an essential and important tool
for constraining any form of BSM physics. In the nmUED model, corrections to Peskin-Takeuchi
parameters S, T, and U appear via the correction to the Fermi constant $G_F$ at tree level. This is a remarkable contrast with respect to the minimal version of the UED model where these corrections appear via one loop processes. Detail study on EWPT for the present version of nmUED model has been provided in \cite{Datta:2015aka, Biswas:2017vhc}. Following the same approach given in refs. \cite{Datta:2015aka, Biswas:2017vhc} we have applied EWPT as one of the constraints in our analysis.  

\end{itemize}
To this end, we would like to mention the range of values of BLT parameters used in our analysis. In general BLT parameters may be positive or negative. However, it is readily evident from Eq.\;\ref{norm} that, for ${r_f}/{R}=-\pi$ the zero-mode solution becomes divergent and beyond  ${r_f}/{R} = - \pi$ the zero-mode fields become ghost-like.  
Hence, any values of BLT parameters lesser than $- \pi R$ should be discarded. Although, for the sake of completeness we have shown numerical results for some negative BLT parameters. However, analysis of electroweak precision data \cite{Datta:2015aka, Biswas:2017vhc} disfavours large portion of negative BLT parameters.
\vspace*{-0.5cm}
\subsection{Numerical results}\label{nr}
We are in a position, where, we would like to present the primary results of our analysis.
\vspace*{-0.5cm}
\subsubsection{Branching ratio}\label{br}
\begin{figure}[ht!]
\begin{center}
\includegraphics[scale=0.9,angle=0]{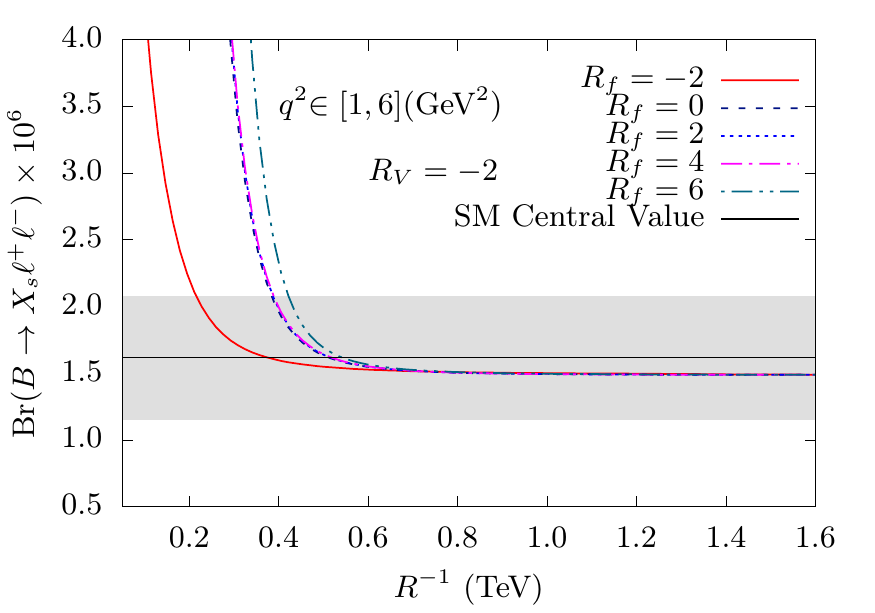}
\includegraphics[scale=0.9,angle=0]{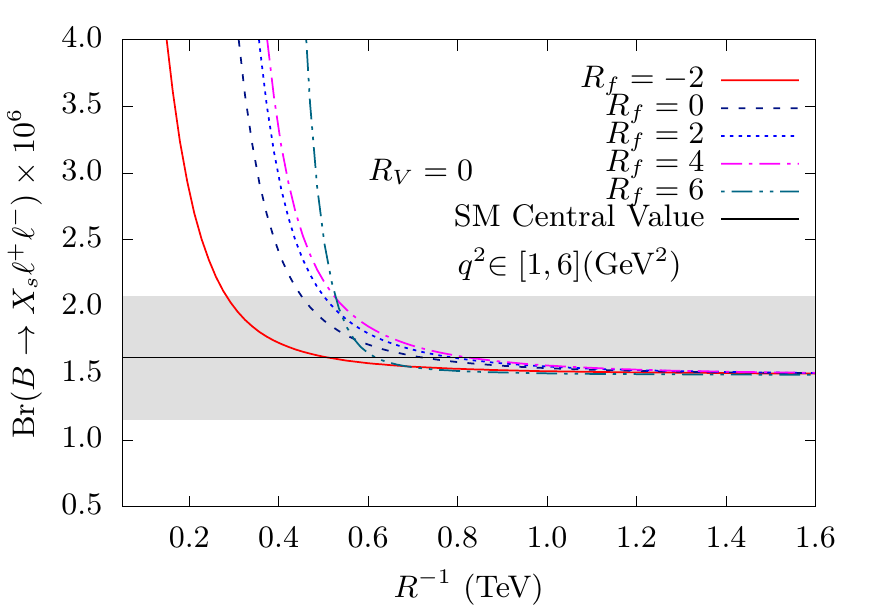}
\includegraphics[scale=0.9,angle=0]{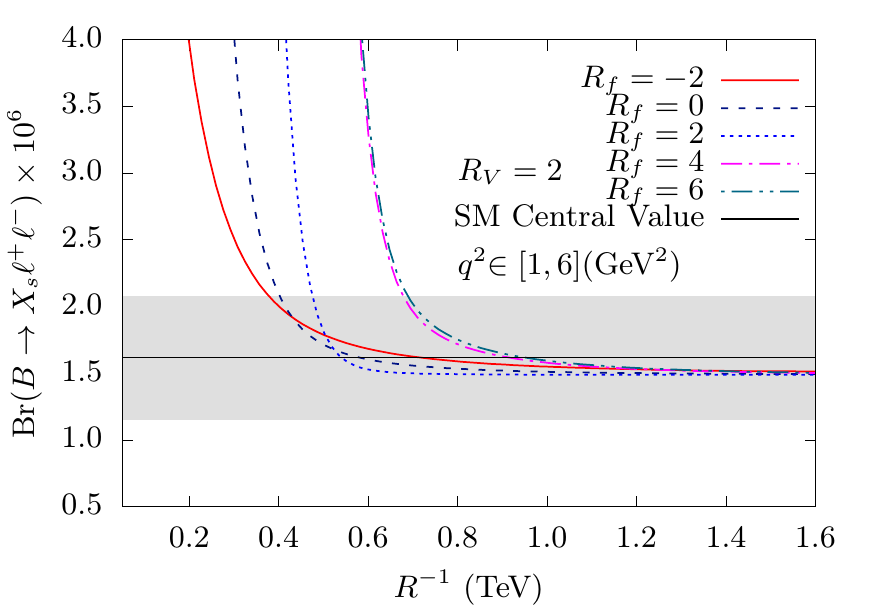}
\includegraphics[scale=0.9,angle=0]{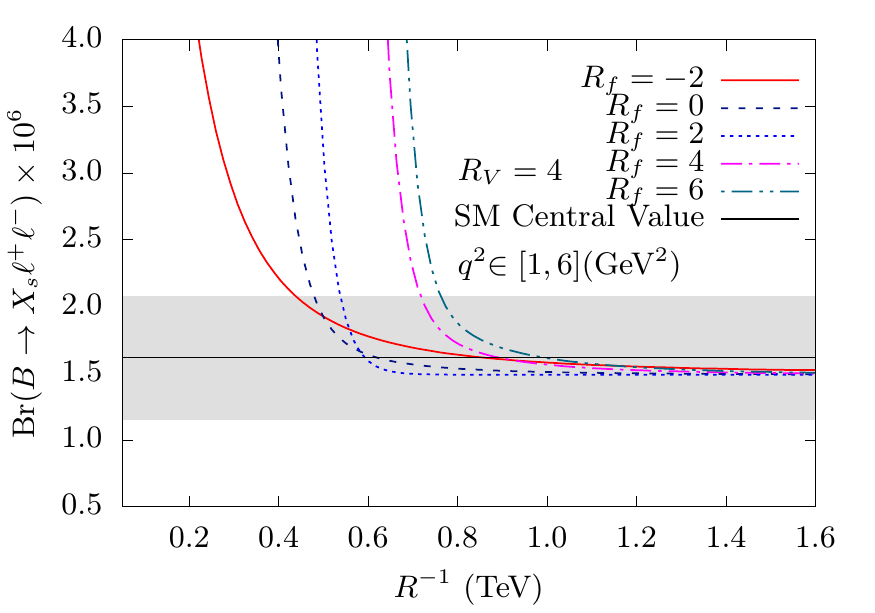}
\includegraphics[scale=0.9,angle=0]{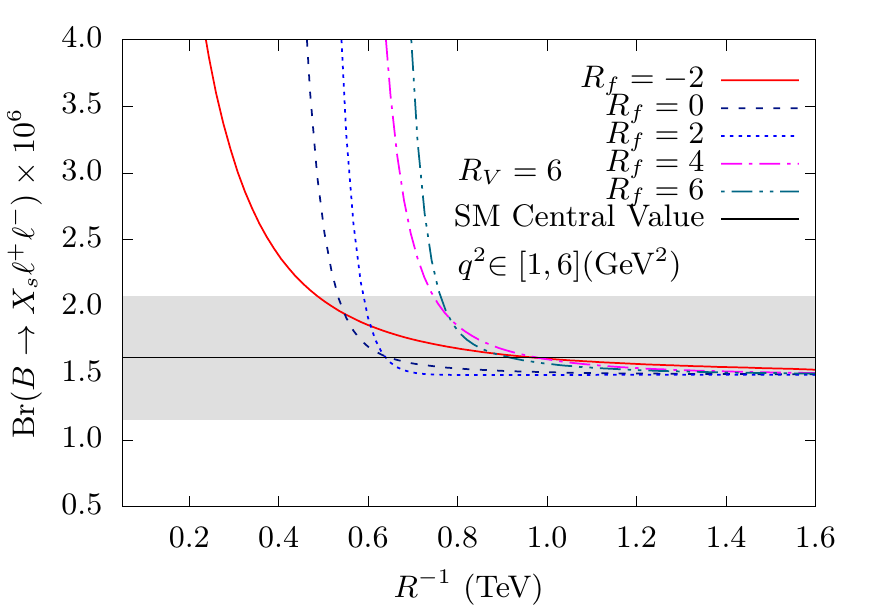}
\caption{Variation of the branching ratio of $B\rightarrow X_s\ell^+\ell^-$ with $R^{-1}$  (TeV) for various values of $R_f(=r_f/R)$. The five panels represent different values of $R_V(=r_V/R)$. We sum the contributions up to 5 KK-levels in different loop functions while calculating WCs. The horizontal grey band depicts the 1$\sigma$ allowed range of experimental value of the branching ratio for $q^2\in [1, 6]~{\rm GeV^2}$.}
\label{low_bran}
\end{center}
\end{figure}

\begin{figure}[ht!]
\begin{center}
\includegraphics[scale=0.9,angle=0]{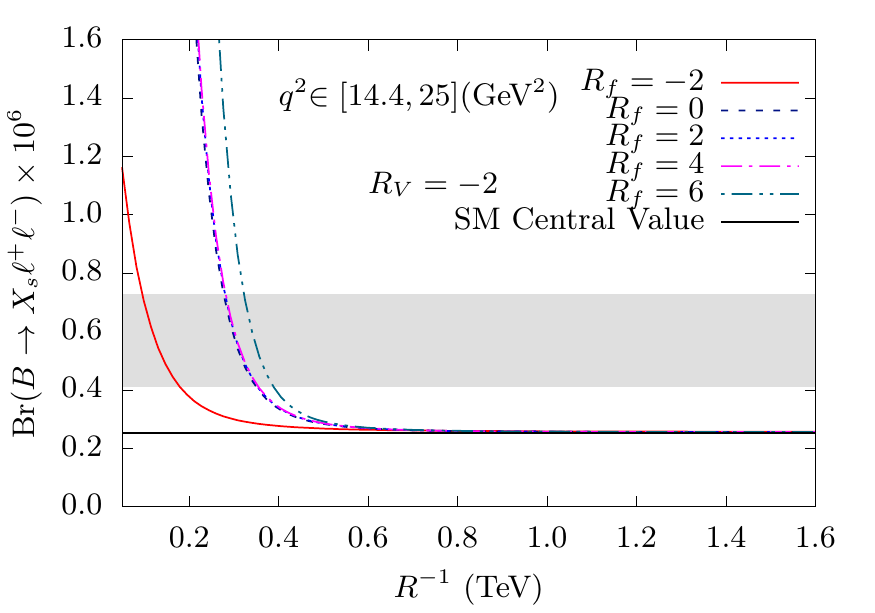}
\includegraphics[scale=0.9,angle=0]{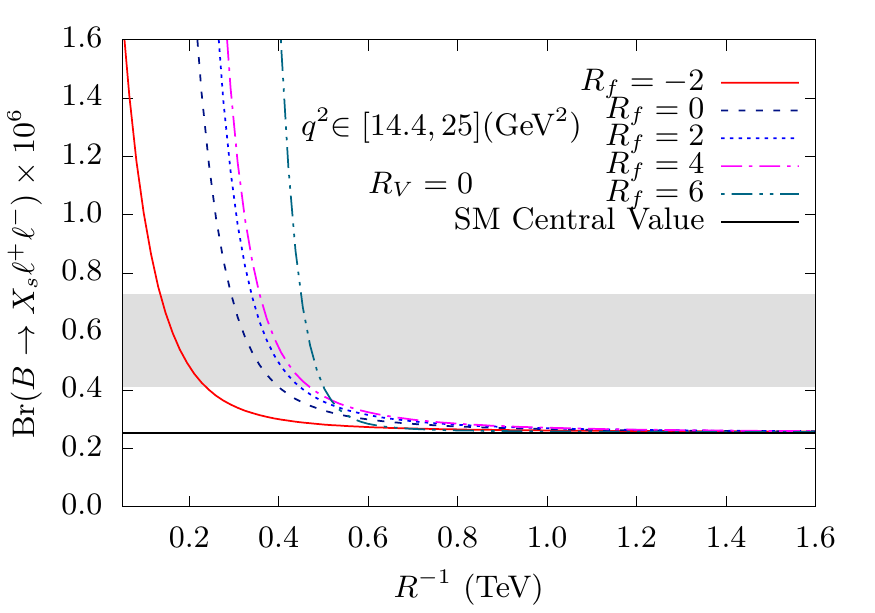}
\includegraphics[scale=0.9,angle=0]{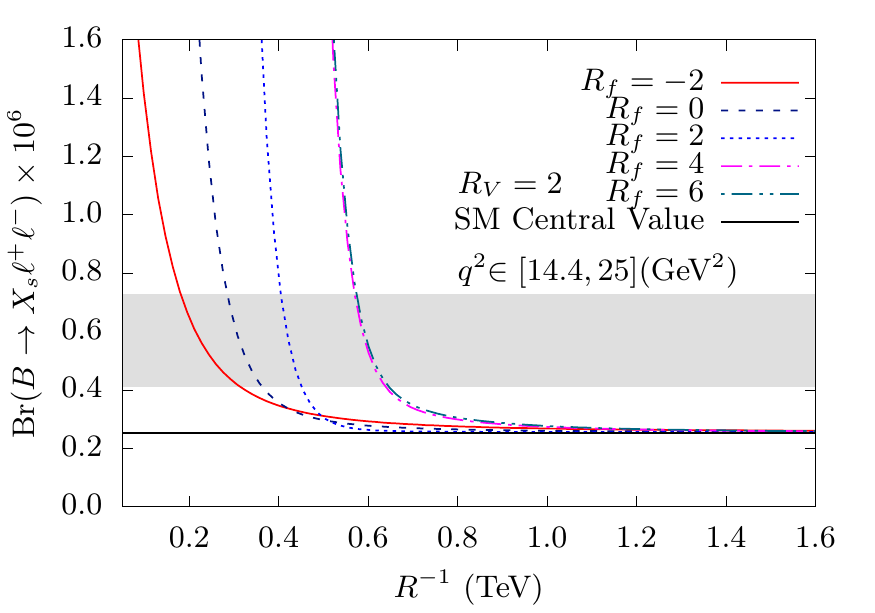}
\includegraphics[scale=0.9,angle=0]{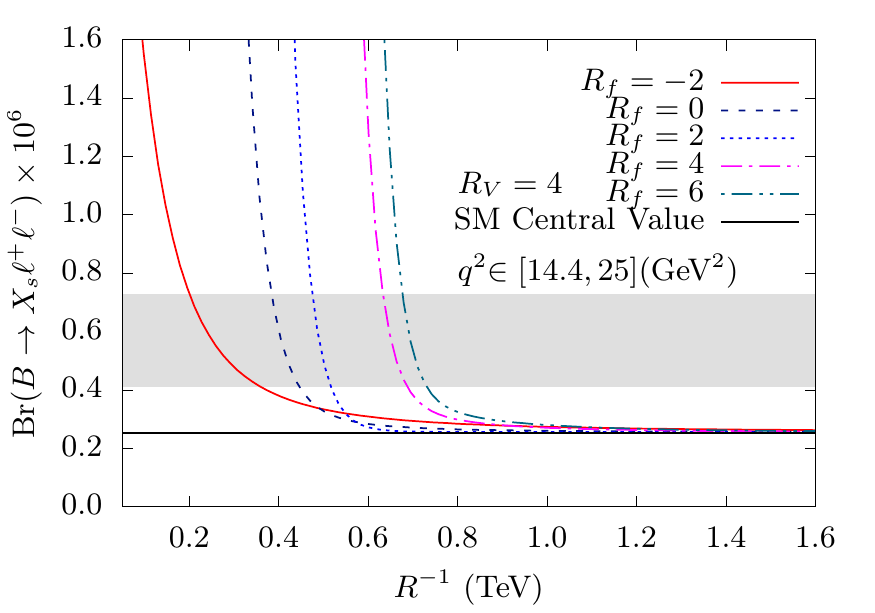}
\includegraphics[scale=0.9,angle=0]{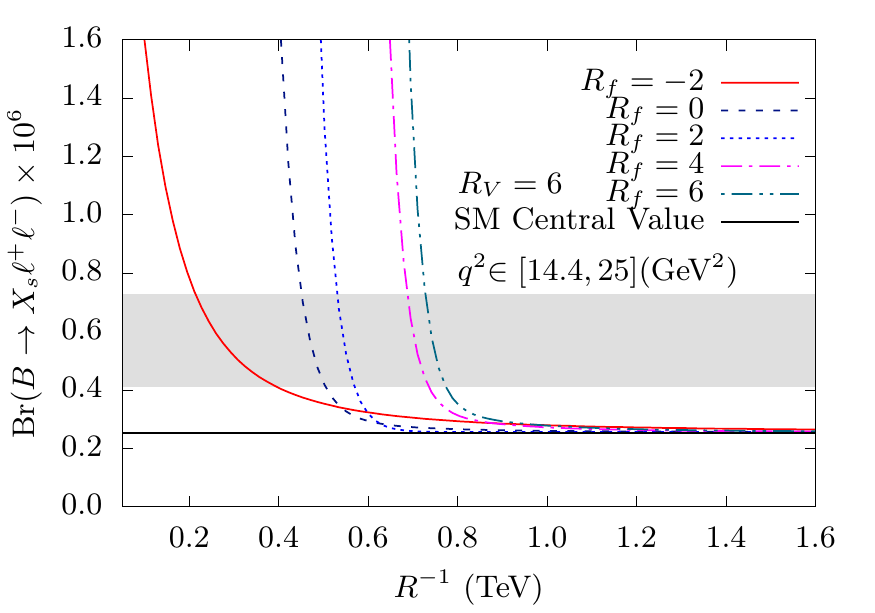}
\caption{Variation of the branching ratio of $B\rightarrow X_s\ell^+\ell^-$ with $R^{-1}$  (TeV) for various values of $R_f(=r_f/R)$. The five panels represent different values of $R_V(=r_V/R)$. We sum the contributions up to 5 KK-levels in different loop functions while calculating WCs. The horizontal grey band depicts the 1$\sigma$ allowed range of experimental value of the branching ratio for $q^2\in [14.4, 25]~{\rm GeV^2}$.}
\label{high_bran}
\end{center}
\end{figure}

In Figs.\;\ref{low_bran} and \ref{high_bran} we have depicted the variation of branching ratio of $B\rightarrow X_s\ell^+\ell^-$ as a function of scaled BLT parameters ($R_V\equiv r_V/R$ and $R_f\equiv r_f/R$) and inverse of the radius of compactification ($R^{-1}$) for two different di-lepton mass square regions $q^2\in [1, 6]~{\rm GeV^2}$ and $q^2\in [14.4, 25]~{\rm GeV^2}$ respectively. We have mentioned earlier  that non-vanishing BLT parameters non-trivially modify the KK-masses and various couplings among the KK-excitations in the nmUED scenario. Therefore, in the following we will discuss that how these BLT parameters affect the concerned decay process. For each of the $q^2$ regions we present five panels corresponding to five different values of scaled gauge BLT parameter $R_V$. In each panel, we show the dependence of the branching ratio with $R^{-1}$ for five different values of scaled fermion BLT parameters $R_f$.

If we focus on a particular curve specified $R_V$ and $R_f$, then we observe that the branching ratio monotonically decreases with respect to increasing values of $R^{-1}$. It is quite expected in a scenario like nmUED, where the masses of KK-excited states are basically characterised by $R^{-1}$, i.e., with the increasing values of $R^{-1}$ the masses of KK-excited states are increased. Therefore with the increasing values of KK-masses, the one loop functions involved in this decay process are suppressed, which in turn decrease the decay width (and branching ratio). Further, depending on the BLT parameters, after a certain value of $R^{-1}$ the branching ratio asymptotically converges to its SM value as $R^{-1}\rightarrow \infty$. This behaviour clearly indicates the decoupling behavior of the KK-mode contribution.

Moreover, it is clearly evident from the Figs.\;\ref{low_bran} and \ref{high_bran} that branching ratio of $B\rightarrow X_s\ell^+\ell^-$  increases with the increment of both of the BLT parameters. For example, if we concentrate on a particular panel specified by a fixed value of $R_V$ then one can see that, with the increasing values of $R_f$ the branching ratio is enhanced. The reason is that, with the increasing values of $R_f$, KK-fermion masses decrease, consequently the loop functions are enhanced. Therefore, the branching ratio increase with higher values of $R_f$. At the same time, if we look at all the panels of any particular figure (either Fig.\;\ref{low_bran} or Fig.\;\ref{high_bran}) then we will readily conclude that the other BLT parameter $R_V$ affects the branching ratio in a similar manner like $R_f$. However, the branching ratio is a bit extra sensitive to the variation of $R_f$ rather than $R_V$. It can be explained by observing the interactions which are involved in this calculation listed in Appendix \ref{fyerul}. As per earlier discussion (see the paragraph before the beginning of the section \ref{sec:Heff:BXsee:nlo}) the interactions are modified by the overlap integrals $I^n_1$ and $I^n_2$. $I^n_1$ modify the interactions of third generations of quarks with charged-Higgs scalar ($H^{(n)\pm}$) and gauge bosons ($W^{(n)\pm}$) while the interactions between the fifth component of $W$-boson and third generations of quarks are modified by $I^n_2$. Therefore, due to the combined effects of the top-Yukawa coupling and $SU(2)$ gauge interaction $I^n_1$ dominates over $I^n_2$ which is controlled by $SU(2)$ gauge interaction only. Hence, $R_f$ has a better control on the $B\rightarrow X_s\ell^+\ell^-$ amplitude ({\it via} $I^n_1$) over $R_V$. 

%\vspace*{-0.5cm}

%\newpage

At this point, we would like to comment on the values of BLT parameters. It is clearly evident from the figures (Figs.\;\ref{low_bran} and \ref{high_bran}) that negative values of BLT parameters are not very encouraging for the present purpose, because we can not get any strong lower limit on $R^{-1}$. For negative BLT parameters the KK-masses are larger with respect to positive BLT parameters. Therefore, enhanced KK-mass suppresses the loop functions, and consequently decay amplitude decreases. Apart from this, constraint of EWPT would prefer larger values of $R^{-1}$ for negative BLT parameters\cite{Datta:2015aka, Biswas:2017vhc}. Hence, in the case of our present purpose the positive values of BLT parameters are more preferable. For example, for $q^2\in [1, 6]~{\rm GeV^2}$ if we choose $R_V = 2, \; R_f = 6$, $R^{-1} > 680 (690) \rm\;GeV$ (see Table \ref{sum-kk_low}) when we consider the sum up to 5(20) KK-levels. On the other hand lower limit on $R^{-1}$ changes to $ > 760 (770)  \rm\;GeV$ for $R_f = R_V = 6$ (see Table \ref{sum-kk_low}). In the case of other region of $q^2 (\in [14.4, 25]~{\rm GeV^2})$, the lower limits on $R^{-1}$ for the above mentioned BLT parameters changes to $ > 570 (580)  \rm\;GeV$ (see Table \ref{sum-kk_high}) and $ > 720 (730)  \rm\;GeV$ (see Table \ref{sum-kk_high}), respectively for the KK-sum up to 5(20) level. We have obtained these limits on $R^{-1}$ by comparing the branching ratio evaluated from the present calculation to the experimental data (given in Eq.\;\ref{EXP-BR-BtoXsll}) with 1$\sigma$ upward error bar. From these numbers we find that the limits are slightly better than that of the results obtained from the analysis $B\rightarrow X_s\gamma$ \cite{Datta:2016flx}, however, in the same ball park of those obtained from the analysis of $B_s \rightarrow \mu^+ \mu^-$ \cite{Datta:2015aka}. Furthermore, if we look at the Figs.\;\ref{low_bran} and \ref{high_bran} (or Tables\;\ref{sum-kk_low} and \ref{sum-kk_high}) then we find that the lower limits on $R^{-1}$ would not drastically change after a certain positive values of BLT parameters. For example, in the present analysis we have restricted ourselves for the choice of BLT parameters (both $R_V$ and $R_f$) up to 6. The reason is that, beyond this choice we expect that the lower limit on $R^{-1}$ would not change significantly for larger values of BLT parameters.

\begin{table}[H]
\begin{center}
\hspace*{-1cm}
\resizebox{19cm}{!}{
\begin{tabular}{|c||c|c||c|c||c|c||c|c||c|c|}
\hline 
{}&\multicolumn{2}{|c||}{$R_V=-2$}&\multicolumn{2}{|c||}{$R_V=0$} &\multicolumn{2}{|c||}{$R_V=2$}&\multicolumn{2}{|c||}{$R_V=4$}&\multicolumn{2}{|c||}{$R_V=6$}\\
\hline{$R_f$}&
5 KK-level &  20 KK-level &
5 KK-level &  20 KK-level &
5 KK-level &  20 KK-level &
5 KK-level &  20 KK-level &
5 KK-level &  20 KK-level \\
\hline
-2&215.73&224.19&283.62&289.23&377.06&381.14&437.53&443.33&487.00&489.26\\
 0&382.15&388.95&451.27&464.93&472.55&482.32&478.76&485.35&530.98&549.54\\
 2&385.45&392.72&498.00&508.18&510.01&518.05&536.48&548.76&588.70&598.28\\
 4&390.26&394.83&525.48&529.81&676.65&688.72&717.88&726.93&745.36&750.21\\
 6&421.04&430.52&528.23&533.45&684.89&694.54&761.85&768.14&764.60&770.42\\
\hline
\end{tabular}
}
\end{center}
\caption[]{Lower limits on $R^{-1}$ (in GeV) evaluated from branching ratio of $B\rightarrow X_s\ell^+\ell^-$ for several values of BLT parameters for $q^2\in [1, 6]~{\rm GeV^2}$ showing the insensitivity on the number of  KK-modes in summation.}
\label{sum-kk_low}
\end{table}

\begin{table}[H]
\begin{center}
\hspace*{-1cm}
\resizebox{19cm}{!}{
\begin{tabular}{|c||c|c||c|c||c|c||c|c||c|c|}
\hline 
{}&\multicolumn{2}{|c||}{$R_V=-2$}&\multicolumn{2}{|c||}{$R_V=0$} &\multicolumn{2}{|c||}{$R_V=2$}&\multicolumn{2}{|c||}{$R_V=4$}&\multicolumn{2}{|c||}{$R_V=6$}\\
\hline{$R_f$}&
5 KK-level &  20 KK-level &
5 KK-level &  20 KK-level &
5 KK-level &  20 KK-level &
5 KK-level &  20 KK-level &
5 KK-level &  20 KK-level \\
\hline
-2&93.98&102.71&135.20&143.26&173.18&186.51&201.16&208.14&214.90&219.42\\
 0&275.38&287.70&294.61&306.26&321.36&335.18&385.31&402.21&451.28&462.56\\
 2&278.12&289.12&335.84&346.81&404.55&415.20&476.01&487.36&528.23&538.32\\
 4&283.62&294.45&357.83&365.52&569.46&566.05&632.68&640.82&687.64&698.48\\
 6&324.84&334.60&451.28&465.75&572.20&586.54&676.65&697.35&726.12&737.88\\
\hline
\end{tabular}
}
\end{center}
\caption[]{Lower limits on $R^{-1}$ (in GeV) evaluated from branching ratio of $B\rightarrow X_s\ell^+\ell^-$ for several values of BLT parameters for $q^2\in [14.4, 25]~{\rm GeV^2}$ showing the insensitivity on the number of  KK-modes in summation.}
\label{sum-kk_high}
\end{table}
%\vspace*{-0.5cm}
In Table \ref{sum-kk_low} and Table \ref{sum-kk_high} (for two different regions of $q^2$) we have enlisted specific values of lower limits on $R^{-1}$ corresponding to different choices of BLT parameters. The numbers in the tables (Table\;\ref{sum-kk_low} and Table\;\ref{sum-kk_high}) also indicate that our results are not very sensitive to the number of KK-levels considered in the sum while calculating loop diagrams corresponding to different WCs. 

\begin{figure}[ht!]
\begin{center}
\includegraphics[scale=0.9,angle=0]{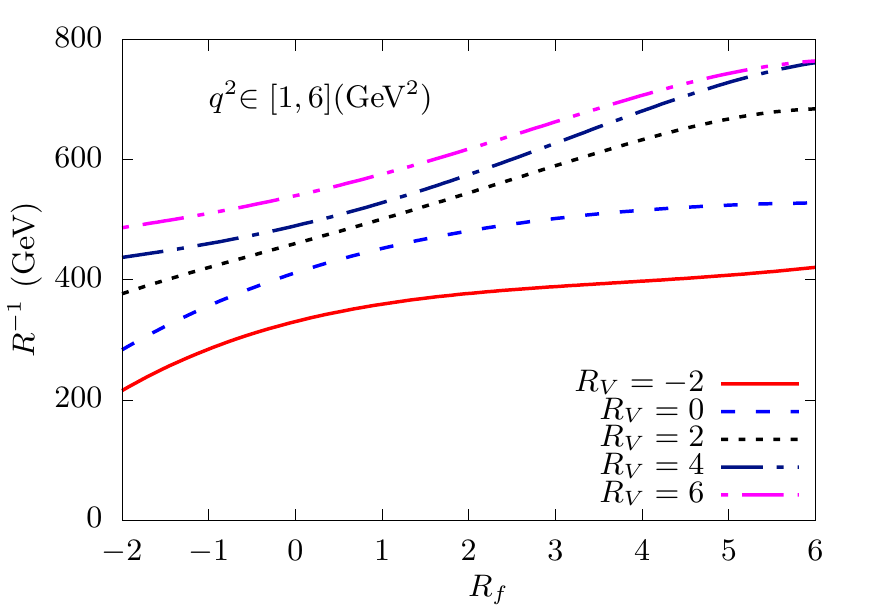}
\includegraphics[scale=0.9,angle=0]{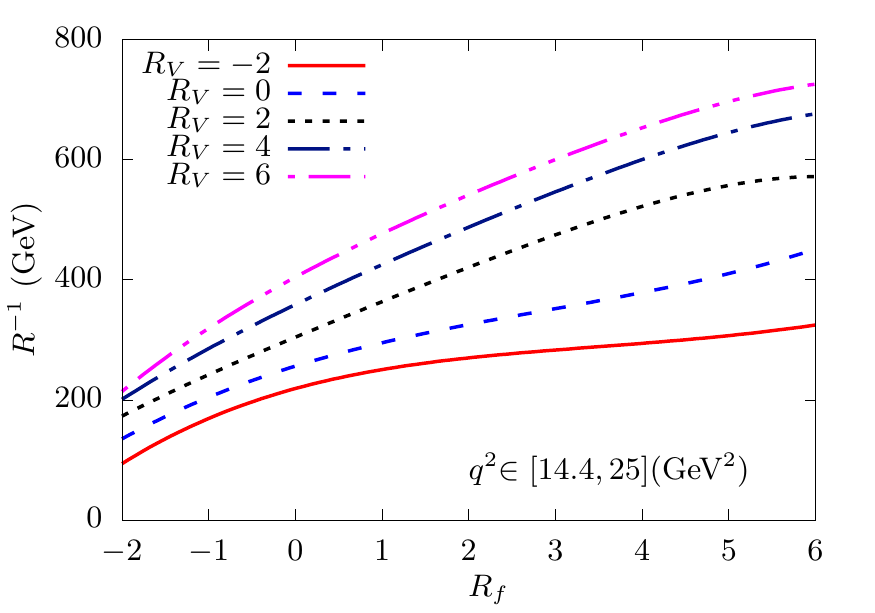}
\caption{Left and right panels represent the exclusion contours obtained from branching ratio of $B\rightarrow X_s\ell^+\ell^-$ decay in $R_f - R^{-1}$ plane for low and high di-lepton mass square regions respectively for five different choices of $R_V$. These exclusion curves have been drawn with the values of lower limit of $R^{-1}$ while we sum the contributions up to 5 KK-levels in different loop functions required for the calculation of WCs. The area below a particular curve (fixed $R_V$) has been excluded by the experimental value of the branching ratio with 1$\sigma$ error bar.}
\label{lowerR}
\end{center}
\end{figure}

In the left and right panels of Fig.\;\ref{lowerR} we present the region of parameter space which has been excluded by the currently measured experimental values of branching ratios of $B\rightarrow X_s\ell^+\ell^-$ for two different $q^2$ regions $[1, 6]~{\rm GeV^2}$ and $[14.4, 25]~{\rm GeV^2}$ respectively. In both of these panels we have depicted contours corresponding to five different values of $R_V$ in $R_f-R^{-1}$ plane. The region under a individual curve (specified by a fixed value of $R_V$)  has been excluded by comparing the experimentally measured branching ratio of $B\rightarrow X_s\ell^+\ell^-$ to its theoretical prediction in the nmUED scenario. The curves represent the contours of constant branching ratios of $B\rightarrow X_s\ell^+\ell^-$ corresponding to the 1$\sigma$ upper limit of its experimentally measured value. One can understand the nature of these contour curves with the help of Figs.\;\ref{low_bran} and \ref{high_bran}. With the larger values of $R^{-1}$ KK-masses increase which lead to suppression in the decay width (and branching ratio). Hence, in order to overcome this suppression one requires larger values of $R_f$ and $R_V$. The larger values of BLT parameter enhance the decay dynamics in two ways. First of all, these would diminish the KK-masses. Secondly, larger values of $R_f$ would increase the interaction strengths via the overlap integral $I^n_1$ where as increasing values of $R_V$ would increase interaction strengths via $I^n_2$. 

To this end, we would like to mention that, as per as the BLT parameters are concerned there is no sharp contrast in behaviour of decay branching ratio between two different regions of $q^2$. However, the lower limits of $R^{-1}$ which we have obtained from our present analysis are slightly different for two different regions of $q^2$. In the case of low $q^2$ region ($\in [1, 6]~{\rm GeV^2}$) the lower limit is higher than that of the case in high $q^2$ region ($\in [14.4, 25]~{\rm GeV^2}$). For example (considering only 5 KK-level sum), in the low $q^2$ region if we set $R_V=4$, $R_f=2$ the lower limit on  $R^{-1}$ is 536.38 GeV while for the same set of BLT parameters $R^{-1}$ is 476.01 GeV for the high $q^2$ region. This feature is true for all combinations of BLT parameters. This feature indicates that, in the second case, masses of the KK-particles which are involved in the loop diagrams are relatively lighter with respect to the first case. This behaviour is quite expected, because in the second case the phase space suppression is larger with respect to first one, hence to compensate this suppression one requires relatively lighter mass particles which are involved in the loop diagrams needed for the calculation of different WCs. 

$\bullet$ {\bf Revisit at the lower limit on {\boldmath$R^{-1}$} obtained from {\boldmath$B\to X_s\ell^+\ell^-$} in UED scenario}\\
Before we proceed any further, we would like to revisit the lower limit on $R^{-1}$ obtained from our analysis in the UED scenario considering the current experimental results of the branching ratio of  $B\rightarrow X_s\ell^+\ell^-$. We can obtain the UED results from our analysis in the limit when both the BLT parameters vanish, i.e., for $R_V=R_f=0$. In this limit KK-mass for $n^{th}$ KK-level simply becomes $nR^{-1}$. Moreover, the overlap integrals $I^n_1$ and $I^n_2$ become unity. Hence, under this circumstance, the functions $D_n(x_t,x_{f^{(n)}},x_{V^{(n)}})$ and $E_n(x_t,x_{f^{(n)}},x_{V^{(n)}})$ given in Eqs.\;\ref{dn} and \ref{en} would transform  themselves into their UED forms. We have explicitly checked that in this vanishing BLT limit the expressions of the functions $D_n(x_t,x_{f^{(n)}},x_{V^{(n)}})$ and $E_n(x_t,x_{f^{(n)}},x_{V^{(n)}})$ are identical with the forms that given in ref.\;\cite{Buras:2003mk}\footnote{The authors of the article \cite{Buras:2003mk} have not considered any radiative corrections to the KK-masses in their analysis. Consequently the KK-mass at the $n^{th}$ KK-level is $nR^{-1}$.}. Under the same vanishing BLT limit condition, similar transformation is also applicable for other functions (e.g., $C_n, Y_n, D'_n$ and $E'_n$ which have been calculated in our previous articles \cite{Datta:2015aka,Datta:2016flx}) required for the present calculations. Now from our present analysis we can readily derive the lower limit on $R^{-1}$ from the Tables \ref{sum-kk_low} and \ref{sum-kk_high}. That is for $R_V = R_f =0$, the value of lower limit on $R^{-1}$ for $q^2\in [1, 6]~{\rm GeV^2}$ is 451.27 GeV, whereas for $q^2\in [14.4, 25]~{\rm GeV^2}$ the value changes to 294.61 GeV. It is needless to say that these results are not very strong but almost consistent with those values that are obtained from previous analyses in UED scenario. For example $(g-2)_\mu$ \cite{Nath:1999aa}, $\rho$-parameter \cite{Appelquist:2002wb}, FCNC process \cite{Buras:2003mk,Buras:2002ej, Agashe:2001xt, Chakraverty:2002qk}, $Zb\bar{b}$ \cite{Jha:2014faa, Oliver:2002up} and electroweak observables \cite{Strumia:1999jm, Rizzo:1999br, Carone:1999nz} put a lower bound of about 300-600 GeV on $R^{-1}$. On the other hand, from the projected tri-lepton signal at 8 TeV LHC one can derive lower limit on $R^{-1}$ up to 1.2 TeV\cite{Belyaev:2012ai, Golling:2016gvc, Gershtein:2013iqa}. At this point it is worth mentioning that the values of lower limit on $R^{-1}$, that obtained from the above mentioned analyses (for minimal version of UED scenario), have already been ruled out by the LHC data. The reason is that the recent analyses including LHC data exclude $R^{-1}$ up to 1.4~TeV~\cite{Choudhury:2016tff, Beuria:2017jez, Chakraborty:2017kjq, Deutschmann:2017bth}.

\subsubsection{Forward-Backward asymmetry}\label{fwbk}
\begin{figure}[t!]
\begin{center}
\includegraphics[scale=0.9,angle=0]{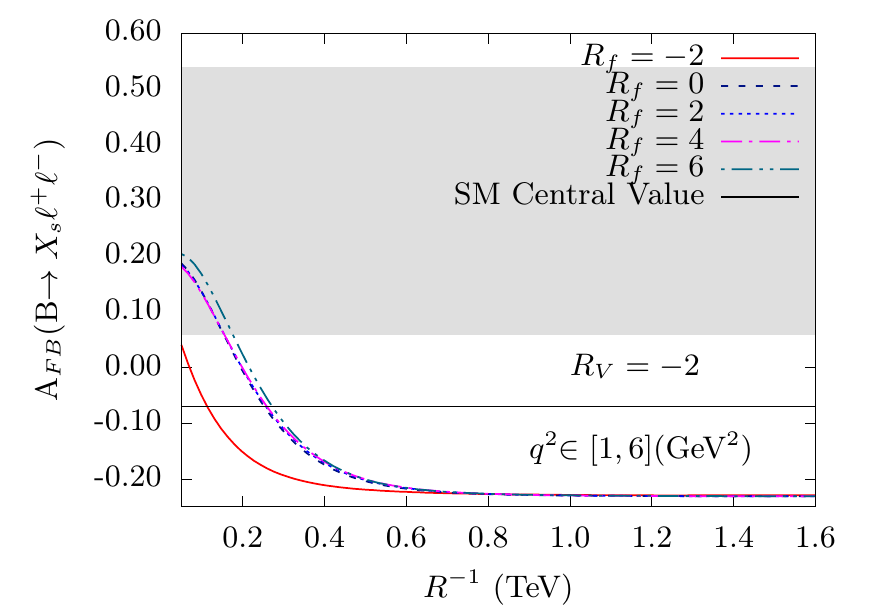}
\includegraphics[scale=0.9,angle=0]{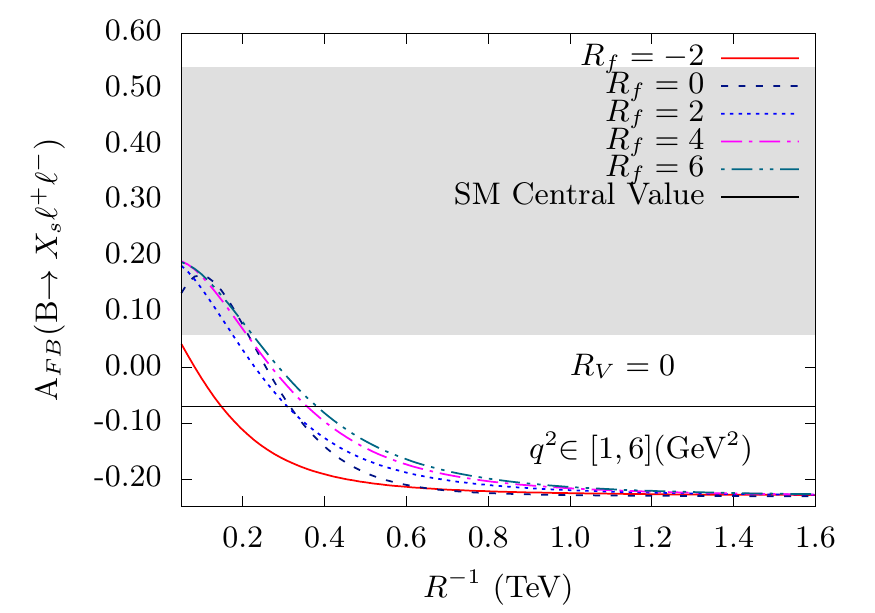}
\includegraphics[scale=0.9,angle=0]{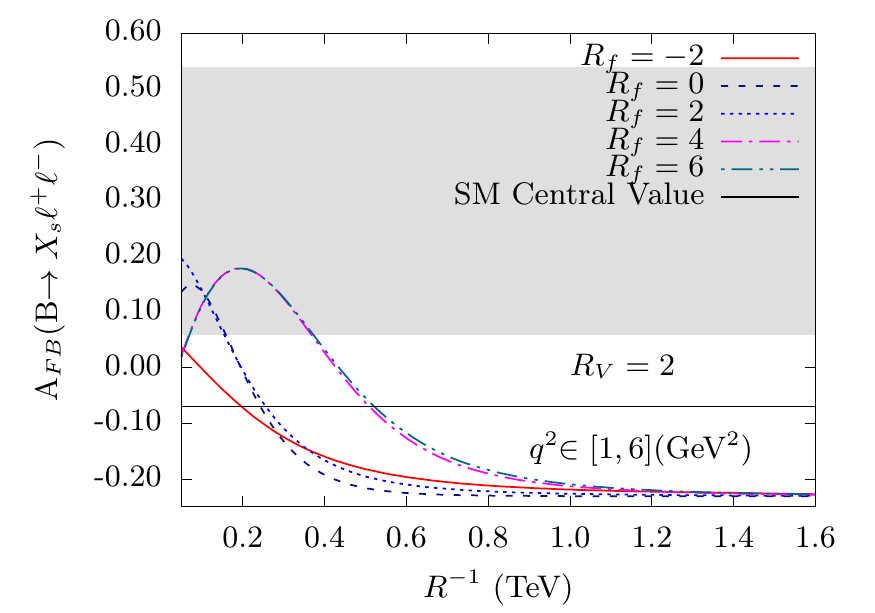}
\includegraphics[scale=0.9,angle=0]{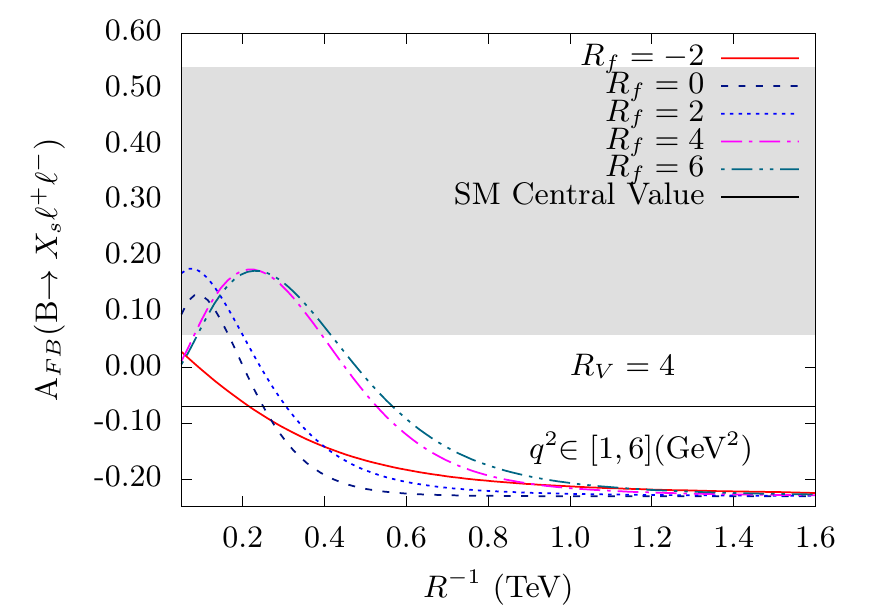}
\includegraphics[scale=0.9,angle=0]{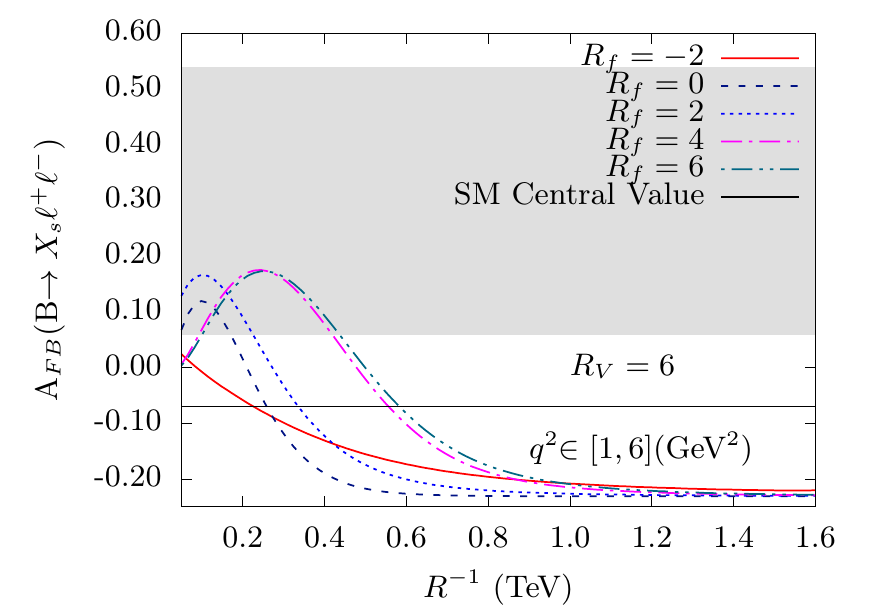}
\caption{Variation of the Forward-Backward asymmetry of $B\rightarrow X_s\ell^+\ell^-$ with $R^{-1}$  (TeV) for various values of $R_f(=r_f/R)$. The five panels represent different values of $R_V(=r_V/R)$. We sum the contributions up to 5 KK-levels in different loop functions while calculating WCs. The horizontal grey band depicts the 1$\sigma$ allowed range of experimental value of the  Forward-Backward asymmetry for $q^2\in [1, 6]~{\rm GeV^2}$.}
\label{low_afb}
\end{center}
\end{figure}

\begin{figure}[t!]
\begin{center}
\includegraphics[scale=0.9,angle=0]{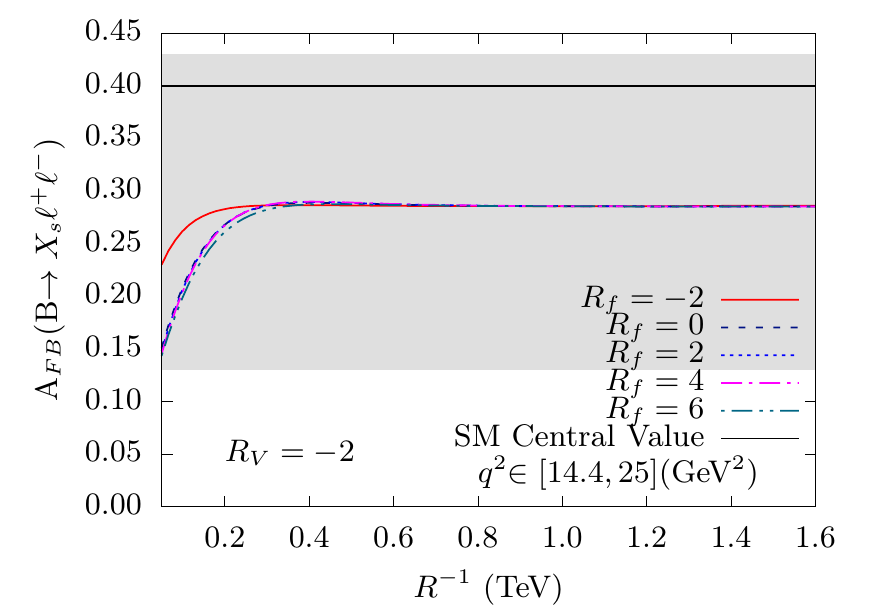}
\includegraphics[scale=0.9,angle=0]{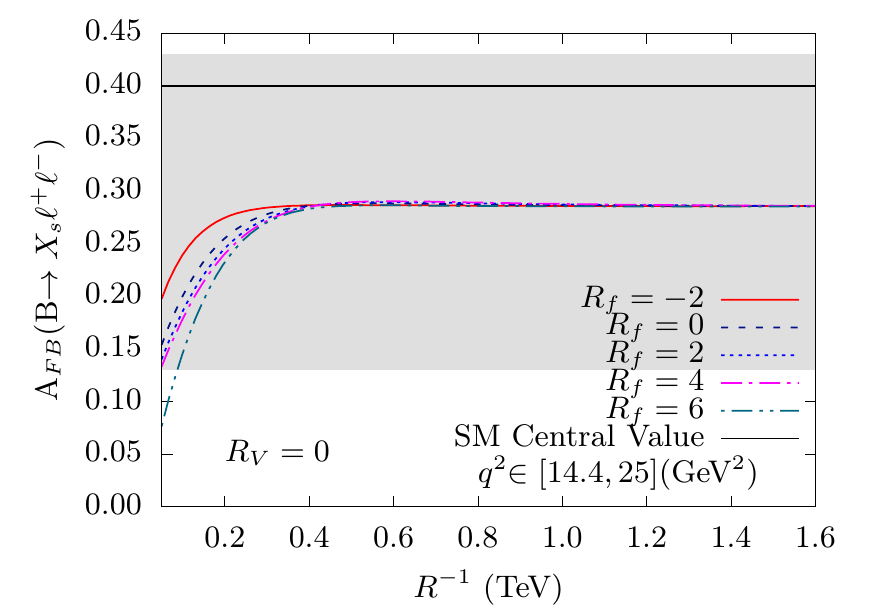}
\includegraphics[scale=0.9,angle=0]{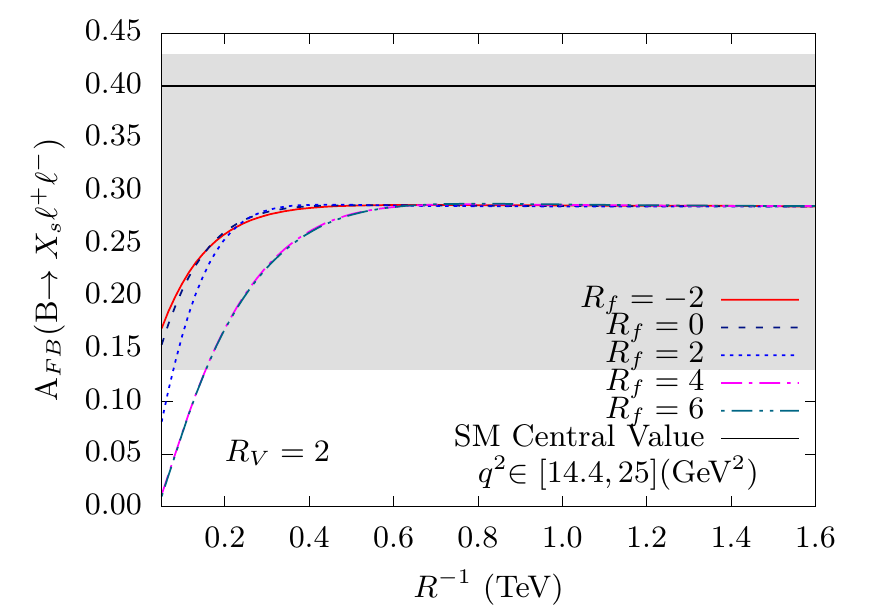}
\includegraphics[scale=0.9,angle=0]{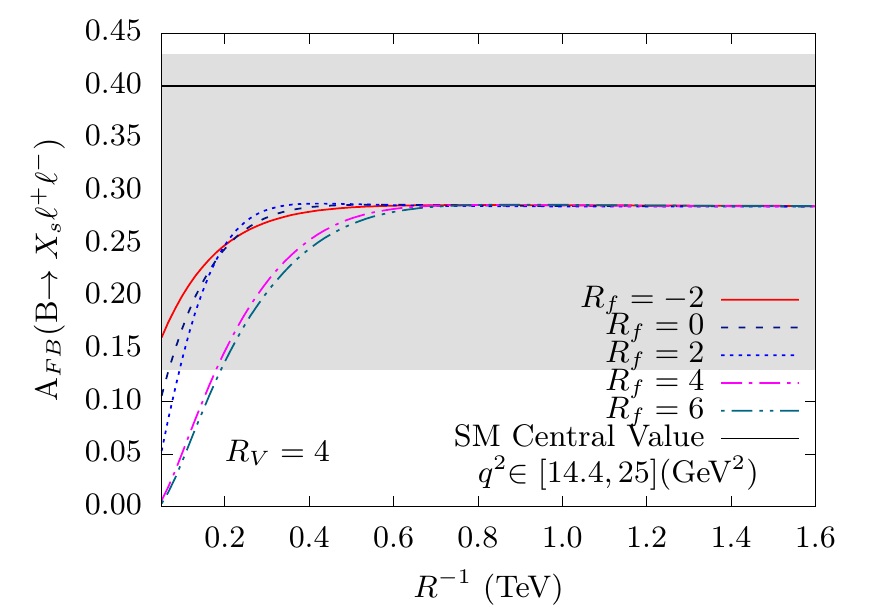}
\includegraphics[scale=0.9,angle=0]{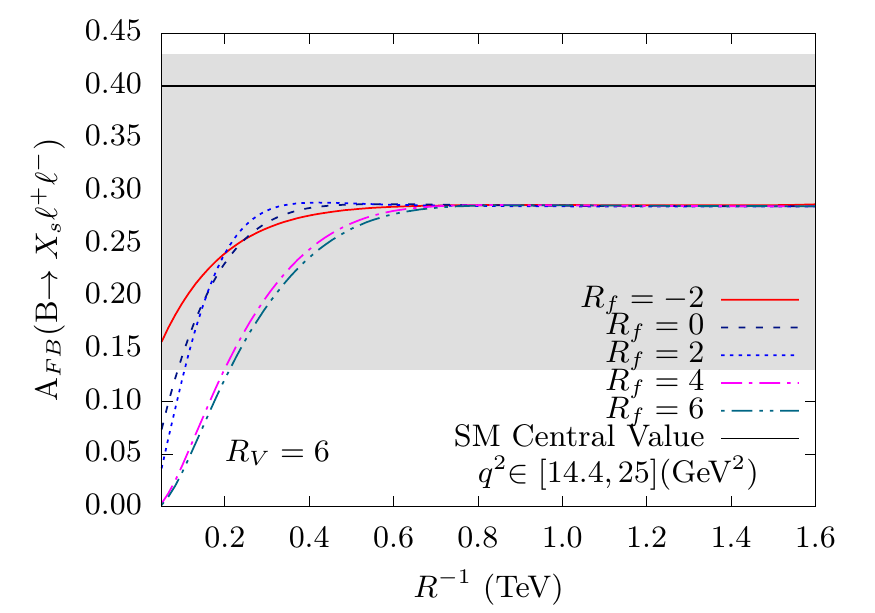}
\caption{Variation of the Forward-Backward asymmetry of $B\rightarrow X_s\ell^+\ell^-$ with $R^{-1}$  (TeV) for various values of $R_f(=r_f/R)$. The five panels represent different values of $R_V(=r_V/R)$. We sum the contributions up to 5 KK-levels in different loop functions while calculating WCs. The horizontal grey band depicts the 1$\sigma$ allowed range of experimental value of the Forward-Backward asymmetry for $q^2\in [14.4, 25]~{\rm GeV^2}$.}
\label{high_afb}
\end{center}
\end{figure}

Finally, in Figs.\;\ref{low_afb} and \ref{high_afb} we have shown the Forward-Backward asymmetry (actually global Forward-Backward asymmetry defined in Eq.\;\ref{global-AFB}) for the decay $B\rightarrow X_s\ell^+\ell^-$ for two $q^2$ regions $[1, 6]~{\rm GeV^2}$ and $[14.4, 25]~{\rm GeV^2}$ respectively. In each figure there are five panels corresponding to five different values of $R_V$. In each panel we have depicted the variation of Forward-Backward asymmetry with respect to $R^{-1}$ for five different values of $R_f$. Unlike the decay branching ratio, the behaviour of Forward-Backward asymmetry has been significantly affected by the two different regions of $q^2$. For example in the high $q^2$ region this asymmetry is always positive for the entire range of given $R^{-1}$ for every combination of BLT parameter, whereas for the low $q^2$ region the sign (either positive or negative) of this asymmetry is crucially dependent on the BLT parameters for the lower values of $R^{-1}$, although, it is always negative for higher values of $R^{-1}$. We have already mentioned that, in the present decay process among all the WCs, only $\Ctilde_{10}$ is moderately enhanced by NP effects. Furthermore, this coefficient is independent of $q^2$ but depends only on the parameters of NP scenario. Now this coefficient has been appeared with a factor proportional to $\frac{q^2}{m^2_b}$ both in the numerator as well as in the denominator of the definition of global Forward-Backward asymmetry. Hence, depending on the value of $q^2$ the factor $\frac{q^2}{m^2_b}$ could play crucial role for the defined asymmetry.

In the case of low $q^2$ region, apart from the factor of $\frac{q^2}{m^2_b}$, some of the WCs could control the behaviour of Forward-Backward asymmetry for the lower values of $R^{-1}$. Since in this situation the masses of KK-modes are not very high, so Forward-Backward asymmetry have been hallmarked by the characteristics of different WCs. Now in every panel specified by a fixed value of $R_V$, we observe that, the asymmetry always shows monotonically decreasing behaviour for negative value of $R_f$. We have earlier mentioned that for negative values of $R_f $ KK-mass high, therefore, the loop functions are suppressed which in turn decrease the asymmetry. On the other hand when $R_f$ changes to positive side then due to relatively smaller values of KK-mass, loop functions are enhanced so that the WCs are increased and consequently the Forward-Backward asymmetry shows increasing behaviour. Then with the increasing values of $R^{-1}$ this asymmetry decreases. Moreover, the same argument is also applicable for the $R_V$, because, if we look at the all panels, then we can readily infer that the above mentioned effects due to $R_f$ are slightly magnified by increasing values of $R_V$. At this point we would like to point out that, using this asymmetry we can maximally achieve the lower limit on $R^{-1}$ up to $\simeq 600$ GeV. This limit can be obtained by comparing the theoretically estimated value of Forward-Backward in the present nmUED model with 1$\sigma$ lower bound of the corresponding experimental data. However, this value is not so competent with one that we have obtained from the branching ratio.
%\newpage
On the other hand for high $q^2$ region, the factor $\frac{q^2}{m^2_b}$ is highly dominating in nature. Therefore, unlike very lower values of $R^{-1}$, the WCs will not get any scope to control the characteristics of the Forward-Backward asymmetry. As a result after a certain value of $R^{-1}$, the numerator as well as the denominator of Forward-Backward asymmetry are totally affected by the same way by the factor $\frac{q^2}{m^2_b}$. Hence, the asymmetry practically becomes independent of $R^{-1}$. This is clearly evident from the plots, where this asymmetry is almost parallel to the $R^{-1}$. Depending on the values of the BLT parameters, the saturation behaviour starts from different values of $R^{-1}$. However, it is also evident from the different panels of Fig.\;\ref{high_afb} that, even for different combination of BLT parameters the threshold points (basically the value of $R^{-1}$) of this saturation behaviour are not very distinct from each other.

\subsubsection{Possible bounds on the nmUED scenario with upcoming measurements by the Belle II for the $B\to X_s \ell^+\ell^-$ observables}
{In near future we will have new measurements by the Belle II experiment for the $B\to X_s \ell^+\ell^-$ observables. Therefore, at this stage it would be very relevant to discuss the possible bounds on the parameter space of the nmUED scenario in light of upcoming measurements by the Belle II experiment for the $B\to X_s \ell^+\ell^-$ observables.  Belle II can significantly improve the present situation with its two orders of magnitude larger data sample. Consequently, we can expect the reduction of systematic uncertainties for the various observables. In order to check the possible bounds on the parameter space of nmUED scenario in the context of upcoming measurements by the Belle II for the $B\to X_s \ell^+\ell^-$ decay observables we follow the prescription given in \cite{Huber:2015sra, Kou:2018nap}. According to this prescription, the bounds can be implemented via the ratios $R_9$ and $R_{10}$ under the assumption of no NP contributions to the electromagnetic and chromomagnetic dipole operators (i.e., $R_{7,8} = 1$), where the ratios are defined as $R_{i}=\frac{C_i}{C^{\rm SM}_i}$ ($C_i$s are different WCs, with $i=7,8,9,10$). In Fig.\;4 of \cite{Huber:2015sra} (in all three panels) we can find a tiny area in $R_9-R_{10}$ plane that could be reached by upcoming results by Belle II experiment. For all cases, within this tiny area both the values of $R_9$ and $R_{10}$ are very much close to the unity. In other words this fact indicates that the deviation between NP and SM prediction is very small. We translate this fact (using lower panel of Fig.\;4 of \cite{Huber:2015sra}) in nmUED scenario in terms of the ratios $R_9$ and $R_{10}$ from which we have obtained the bounds on the model parameters from the perspective of upcoming measurements by the Belle II for the $B\to X_s \ell^+\ell^-$ observables. 

In the nmUED scenario, we have determined the values of the model parameters for which the ratios $R_9$ and $R_{10}$ should be restricted within the tiny area in $R_9-R_{10}$ plane that could be reached by upcoming results by Belle II experiment. The values of the lower limit of $R^{-1}$ for different combination of BLT parameters $R_f$ and $R_V$ have been slightly shifted to the higher values with respect to that of the values which we have obtained from our main analysis of this article. For example, when $R_V=2$ and $R_f=4$, the lower limit on $R^{-1}$ is 680.27 GeV, while this limit changes to 772.81 GeV for $R_V=6$ and $R_f=6$. This behaviour is true for all combination of BLT parameters. Here, we would like to mention that, these values are obtained when we consider the sum up to 5 KK-levels. This kind of result implies that the deviation between the SM expectation and the upcoming measurement by Belle II for the $B\to X_s \ell^+\ell^-$ decay observables will be decreasing in nature. Consequently, the role of NP is expected to be more restricted for the $B\to X_s \ell^+\ell^-$ decay observables. Therefore, one can constrain any NP model more precisely using the  upcoming measurement by Belle II for the $B\to X_s \ell^+\ell^-$ decay observables. Moreover, the tendency of increasing of the lower limit of $R^{-1}$ indicates that NP model (in our case nmUED scenario) approaches to the direction of decoupling limit. Because, we have already mentioned that in a scenario like nmUED, where the masses of KK-excited states (NP particles in the present case) are essentially characterised by $R^{-1}$, therefore, with the increasing values of $R^{-1}$ the masses of KK-excited states are increased. Consequently, the effect of these KK-excited states will be decreased.} 
{%This type of behaviour has also been observed from the perspective of other physical observables and in other NP models. From this phenomena we can roughly infer that proposed NP models are of the decoupling type, with (possibly) highly massive particles and/or very feeble interactions.}

\section{Summary and conclusion}\label{concl}
In view of the findings of new physics effects, we have estimated the contributions of KK-excitations to the decay of $B\rightarrow X_s\ell^+\ell^-$ in a $4 + 1$ dimensional non-minimal Universal Extra Dimensional scenario which is allowed to propagate all Standard Model particles. This specific scenario is characterised by different boundary localised terms (kinetic, Yukawa etc.). Actually in the 5-dimensional Universal Extra Dimensional scenario, the unknown radiative corrections to the masses and couplings are parametrised by the strength of these boundary localised terms. Hence, in the presence of these terms the KK-mass spectra as well as the interaction strengths among the various KK-excitations are transformed in a non-trivial manner in the 4-dimensional effective theory with respect to the minimal version of Universal Extra Dimensional scenario. In the present article we have used two different categories of BLT parameters. For example strengths for the  boundary terms of fermions and Yukawa interactions are represented by $r_f$ while $r_V$ represents the strengths of boundary terms for the gauge as well as Higgs sectors. We have examined the effects of these BLT parameters on $B\rightarrow X_s\ell^+\ell^-$ decay process.   

The effective Hamiltonian for the decay process $B\rightarrow X_s\ell^+\ell^-$ is characterised by several Wilson Coefficients $C_7$, $C_9$ and $C_{10}$. {In non-minimal Universal Extra Dimensional scenario the coefficients $C_7$ and $C_{10}$ have already been calculated in our previous articles. However, for the first time we have calculated the coefficient $C_9$ in the non-minimal Universal Extra Dimensional scenario using the relevant Feynman (penguin) diagrams shown in Fig.\;\ref{magnetic_pen}. With these several Wilson Coefficients we have computed the coefficients of electroweak dipole operators for photon and gluon for the first time in the non-minimal Universal Extra Dimensional scenario}. Applying the advantage of Glashow Iliopoulos Maiani (GIM) mechanism we have included contributions from all three generations of quarks in our analysis. We evaluate the total contribution that obtained from the penguin diagrams and then added it with the corresponding Standard Model counterpart. Considering a recent analysis relating the stability on Higgs boson mass and cut-off of a Universal Extra Dimensional scenario \cite{Datta:2012db}, we have considered the summation up to 5 KK-levels in our calculation. Furthermore, we have incorporated next-to-leading QCD corrections in our analysis. 

For the present decay process in order to maintain preturbativity, one has to impose appropriate choice of kinematic cuts to eliminate $c\bar{c}$ resonances which shows large peaks in the di-lepton invariant mass spectrum. Consequently, this gives two distinct perturbative di-lepton invariant mass square regions, called the low di-lepton mass square region $1~{\rm GeV^2} < q^2 < 6~{\rm GeV^2}$, and also the high di-lepton mass square region with $q^2 > 14.4~{\rm GeV^2}$. In these two regions, experimental data for branching ratio as well as Forward-Backward asymmetry are available for the decay $B\rightarrow X_s \ell^+\ell^-$. However, there exists only a narrow window between the Standard Model prediction and the experimental data for both the regions and for both quantities (branching ratio and Forward-Backward asymmetry). Comparing our theoretical predictions with the corresponding experimental data (with 1$\sigma$ error bar) we have constrained the parameter space of the present version of non-minimal Universal Extra Dimensional scenario. During our analysis we have used the branching ratios of some rare decay processes such as $B_s\to \mu^+\mu^-$ and $B\to X_s\gamma$ as well as electroweak precision data as constraints. 

As we have already mentioned that from our analysis we can also reproduce the results of the minimal version of Universal Extra Dimensional scenario by setting the BLT parameters as zero (i.e., $R_f = R_V = 0$). Hence, from our analysis we have revisited the lower limit on $R^{-1}$ in the framework of minimal Universal Extra Dimensional scenario. Using the experimental data of the branching ratio the lower limit becomes 451.27 (294.61) GeV for low (high) $q^2$ region. Definitely these results are comparable with those values that are obtained from the earlier analysis exist in the literature, although, ruled out from recent collider analysis at the LHC. However, by the virtue of the presence of different non-zero BLT parameters we can improve the results of lower limit on $R^{-1}$ in the present version of non-minimal Universal Extra Dimensional scenario. For example, for $R_V=6$ and $R_f=6$ using branching ratio we obtain the lower limit of $R^{-1}$ $\geq 760$ GeV for the low $q^2$ region while the limit changes to $R^{-1}$ $\geq 720$ for high $q^2$ region. Obviously these results in the context of non-minimal Universal Extra Dimensional scenario is very promising because it excludes a large portion of the parameters space of the present scenario. Also the obtained lower limit on $R^{-1}$ is in the same ball park as the limit obtained from previous analysis on $B_s \to \mu^+\mu^-$ \cite{Datta:2015aka} in non-minimal Universal Extra Dimensional scenario. Furthermore, from Fig.\;\ref{lowerR} it is clearly evident that the lower limits on $R^{-1}$ are relatively more competitive for positive values of the BLT parameters rather than their negative values. Unfortunately, the limits which we have obtained on the parameters space (of non-minimal Universal Extra Dimensional scenario) using the Forward-Backward asymmetry of the decay $B\rightarrow X_s\ell^+\ell^-$ are not so competitive.

{Moreover, we have tried to determine the possible bounds on the model parameters of non-minimal Universal Extra Dimensional scenario with upcoming measurements by the Belle II for the $B\to X_s \ell^+\ell^-$ observables. We have found that, for all combination of BLT parameters $R_f$ and $R_V$ the lower limit of $R^{-1}$ have been slightly shifted to the higher values with respect to that of the values which we have achieved from our main analysis of this article.}

%\newpage
{\bf Acknowledgements} The author is grateful to Anindya Datta for taking part at the
initial stage and for many useful discussions. The author is very thankful to Andrzej J. Buras for useful suggestion. 
The author would also like to give thank to Anirban Biswas for computational support. 
%\newpage
%\vspace*{-2cm}
\begin{appendices}
\renewcommand{\thesection}{\Alph{section}}
\renewcommand{\theequation}{\thesection-\arabic{equation}} 

\setcounter{equation}{0}  
\section{Some important functions and Wilson Coefficients that are required for the calculation of \boldmath$B\to X_s \ell^+\ell^-$ in nmUED}\label{NDR}
\begin{itemize}
\item Functions\cite{Buras:1994dj}:
\begin{eqnarray}
%%%%%%%%%%%%%%%%%%%%%%%%%%%%%%%%%%%%%%%%%%%%%%%%%%%
\omega\left(\frac{q^2}{m^2_b}\right)&=&-{2\over9}\pi^2-{4\over3}{\rm Li}_{_2}\left(\frac{q^2}{m^2_b}\right)-{2\over3}\ln \left(\frac{q^2}{m^2_b}\right)\ln\left(1-\frac{q^2}{m^2_b}\right) \\ \nonumber 
&&-{5+4\frac{q^2}{m^2_b}\over3\bigg(1+2\frac{q^2}{m^2_b}\bigg)}\ln\left(1-\frac{q^2}{m^2_b}\right) -{2\frac{q^2}{m^2_b}\bigg(1+\frac{q^2}{m^2_b}\bigg)\bigg(1-2\frac{q^2}{m^2_b}\bigg)\over3\bigg(1-\frac{q^2}{m^2_b}\bigg)^2\bigg(1+2\frac{q^2}{m^2_b}\bigg)}\ln \left(\frac{q^2}{m^2_b}\right) \\ \nonumber 
&&+{5+9\frac{q^2}{m^2_b}-6\left(\frac{q^2}{m^2_b}\right)^2\over6\bigg(1-\frac{q^2}{m^2_b}\bigg)\bigg(1+2\frac{q^2}{m^2_b}\bigg)}\;,
\end{eqnarray}
and
\begin{eqnarray}
\label{hz}
h\left(z,\frac{q^2}{m^2_b}\right)&=&{8\over27}-{8\over9}\ln{m_{_b}\over\mu}-{8\over9}\ln z+{16z^2m^2_b\over9q^2} \\ \nonumber 
&&-{4\over9}\left(1+{2z^2m^2_b\over q^2}\right)\sqrt{\bigg|1-{4z^2m^2_b\over q^2}\bigg|}
\left\{\begin{array}{ll}\ln\Bigg|{\sqrt{1-\frac{4z^2m^2_b}{q^2}}+1\over\sqrt{1-\frac{4z^2m^2_b}{q^2}}-1}\Bigg|-i\pi,
&{\rm if}\: \frac{4z^2m^2_b}{q^2}<1\\
2\arctan{1\over\sqrt{\frac{4z^2m^2_b}{q^2}-1}},&{\rm if}\: \frac{4z^2m^2_b}{q^2}>1\end{array}\right. \\
\label{h0}
h\left(0,\frac{q^2}{m^2_b}\right)&=&{8\over27}-{8\over9}\ln{m_{_b}\over\mu}-{4\over9}\ln \left(\frac{q^2}{m^2_b}\right) + \frac 49 i\pi.
\end{eqnarray}
\newpage
\item Wilson Coefficients:\\{\underline{$C_1\ldots C_6$}}\cite{Buras:1994qa}
\begin{eqnarray}
\label{c1}
C^{(0)}_1(M_W)&=&\frac{11}{2}\frac{\alpha_s(M_W)}{4\pi}\;, \\
\label{c2}
C^{(0)}_2(M_W)&=&1-\frac{11}{6}\frac{\alpha_s(M_W)}{4\pi}\;,\\
\label{c3_4}
C^{(0)}_3(M_W)&=&-\frac 13 C^{(0)}_4(M_W)=-\frac{\alpha_s(M_W)}{24\pi}=\widetilde{E}(x_t, r_f, r_V, R^{-1})\;,\\
C^{(0)}_5(M_W)&=&-\frac 13 C^{(0)}_6(M_W)=-\frac{\alpha_s(M_W)}{24\pi}=\widetilde{E}(x_t, r_f, r_V, R^{-1})\;,
\label{c5_5}
\end{eqnarray}
where,
\begin{eqnarray}
\label{Etilde}
\widetilde{E}(x_t, r_f, r_V, R^{-1})=E(x_t, r_f, r_V, R^{-1})-\frac 23\;.
\end{eqnarray}
{\underline{$C_7$}}\cite{Buras:1994dj}
\begin{eqnarray}
\label{C7eff}
C_{7\gamma}^{(0){\rm eff}} & = & 
\eta^\frac{16}{23} C_{7\gamma}^{(0)}(M_W) + \frac{8}{3}
\left(\eta^\frac{14}{23} - \eta^\frac{16}{23}\right) C_{8G}^{(0)}(M_W) +
 C_2^{(0)}(M_W)\sum_{i=1}^8 h_i \eta^{a_i},
%\\
%\label{C8eff}
%C_{8G}^{(0)eff}(\mu_b) & = & 
%\eta^\frac{14}{23} C_{8G}^{(0)}(\mu_W) 
%   + C_2^{(0)}(\mu_W) \sum_{i=1}^8 \bar h_i \eta^{a_i},
\end{eqnarray}
with
\begin{equation}
\eta  =  \frac{\alpha_s(M_W)}{\alpha_s(m_b)},~~~\alpha_s(m_b) = \frac{\alpha_s(M_Z)}{1 
- \beta_0 \frac{\alpha_s(M_Z)}{2\pi} \, \ln(M_Z/m_b)}, \qquad 
\beta_0=\frac{23}{3}~,
\label{eq:asmumz}
\end{equation}
and 
\begin{eqnarray}\label{c78}
%C^{(0)}_2(M_W) &=& 1,\\                             
C^{(0)}_{7\gamma} (M_W) &=& -\frac{1}{2} D'(x_t, r_f, r_V, R^{-1}),\\ 
C^{(0)}_{8G}(M_W) &=& -\frac{1}{2} E'(x_t, r_f, r_V, R^{-1}).
\end{eqnarray}
The values of $a_i$, $h_i$ and $\bar h_i$ can be obtained from \cite{Buras:2003mk}. The functions $D'(x_t, r_f, r_V, R^{-1})$ and $E'(x_t, r_f, r_V, R^{-1})$ are the total (SM+nmUED) contributions at the LO as given in \cite{Datta:2016flx}.
\end{itemize}

\section{Feynman rules for \boldmath{$B\rightarrow X_s\ell^+\ell^-$} in nmUED}\label{fyerul}
In this Appendix we have given the relevant Feynman rules for our calculations. All momenta and fields are assumed to be incoming. $\hat{A}$ represents background photon field.
\newpage
1) $\hat{A}^{\mu}W^{\nu\pm}S^{\mp}$
$\displaystyle : {g_2}{s_w M_{W^{(n)}}} g_{\mu\nu} C$, where $C$ is given in the following:

\begin{equation}
 \begin{aligned}
  \hat{A}^{\mu} W^{\nu(n)+} G^{(n)-}: C   &= 0,\\ 
  \hat{A}^{\mu} W^{\nu(n)-} G^{(n)+}: C   &= 0,\\
  \hat{A}^{\mu} W^{\nu(n)+} H^{(n)-}: C   &= 0,\\
  \hat{A}^{\mu} W^{\nu(n)-} H^{(n)+}: C   &= 0,  
 \end{aligned}
\end{equation}
where $g_2$ is represent the $SU(2)$ gauge coupling constant while $s_w$ is denoted as $\sin$ of Weinberg angle ($\theta_w$).

%\newpage
2) $\hat{A}^{\mu}S^{\pm}_1S^{\mp}_2$
$\displaystyle : -{ig_2}{s_w} (k_2-k_1)_{\mu} C$, where $C$ is given in the following:
%\end{minipage}

\begin{equation}
 \begin{aligned}
  \hat{A}^{\mu} G^{(n)+} G^{(n)-}: C &= 1,\\   
  \hat{A}^{\mu} H^{(n)+} H^{(n)-}: C &= 1,\\
  \hat{A}^{\mu} G^{(n)+} H^{(n)-}: C &= 0,\\ 
  \hat{A}^{\mu} G^{(n)-} H^{(n)+}: C &= 0,
\end{aligned}
\end{equation}
where the scalar fields $S\equiv H,G.$

3) $\hat{A}^{\mu}(k_1)W^{\nu+}(k_2)W^{\lambda-}(k_3)$
$\displaystyle :$
\begin{equation}
 ig_2s_w \left[g_{\mu\nu} (k_2
      -k_1+k_3)_\lambda + g_{\mu\lambda} (k_1 -k_3 -
      k_2)_\nu
  + g_{\lambda\nu} (k_3 -k_2)_\mu\right].
\end{equation}

%\newpage
4) $\hat{A}^{\mu}{\overline{f}_1} f_2$
  $\displaystyle  : {i g_2}{s_w} \gamma_\mu C$, where $C$ is given in the following:

\begin{equation}
 \begin{aligned}
  \hat{A}^{\mu} \bar{u_i} u_i: C &= \frac23,\\   
  \hat{A}^{\mu} {\overline{T}^{1(n)}_i} T^{1(n)}_i: C &= \frac23,\\
  \hat{A}^{\mu} {\overline{T}^{2(n)}_i} T^{2(n)}_i: C &= \frac23,\\ 
  \hat{A}^{\mu} {\overline{T}^{1(n)}_i} T^{2(n)}_i: C &= 0,\\
  \hat{A}^{\mu} {\overline{T}^{2(n)}_i} T^{2(n)}_i: C &= 0.
\end{aligned}
\end{equation}

\newpage
5) $G^{\mu}{\overline{f}_1} f_2$
  $\displaystyle  : {i g_s}{T^a_{\alpha\beta}} \gamma_\mu C$, where $C$ is given in the following:

\begin{equation}
 \begin{aligned}
  G^{\mu} \bar{u_i} u_i: C &= 1,\\   
  G^{\mu} {\overline{T}^{1(n)}_i} T^{1(n)}_i: C &= 1,\\
  G^{\mu} {\overline{T}^{2(n)}_i} T^{2(n)}_i: C &= 1,\\ 
  G^{\mu} {\overline{T}^{1(n)}_i} T^{2(n)}_i: C &= 0,\\
  G^{\mu} {\overline{T}^{2(n)}_i} T^{2(n)}_i: C &= 0.
\end{aligned}
\end{equation}

6) $S^{\pm}{\overline{f}_1} f_2$
  $\displaystyle  = \frac{g_2}{\sqrt{2} M_{W^{(n)}}} (P_L C_L + P_R C_R)$, where $C_L$ and $C_R$ are given in the following:
%\end{minipage}

%\begin{alignat}{4}
\begin{equation}
\begin{aligned}
  & G^+ \bar{u_i} d_j :  &
  &\left\{\begin{array}{l}C_L = -m_i V_{ij},\\
      C_R = m_j V_{ij},\end{array}\right.
  &&G^- \bar{d_j} u_i :     &
  &\left\{\begin{array}{l}C_L = -m_j V_{ij}^*,\\
      C_R = m_i V_{ij}^*,\end{array}\right.\\
  & G^{(n)+}{\overline{T}^{1(n)}_i} d_j :  &
  &\left\{\begin{array}{l}C_L = -m_1^{(i)} V_{ij},\\
      C_R = M_1^{(i,j)} V_{ij},\end{array}\right.
  &&G^{(n)-}\bar{d_j}T^{1(n)}_i :   &
  &\left\{\begin{array}{l}C_L = -M_1^{(i,j)} V_{ij}^*,\\
     C_R = m_1^{(i)} V_{ij}^*,\end{array}\right.\\
  & G^{(n)+}{\overline{T}^{2(n)}_i} d_j :  &
  &\left\{\begin{array}{l}C_L = m_2^{(i)} V_{ij},\\
      C_R =-M_2^{(i,j)} V_{ij},\end{array}\right.
  &&G^{(n)-}\bar{d_j}T^{2(n)}_i :   &
  &\left\{\begin{array}{l}C_L = M_2^{(i,j)} V_{ij}^*,\\
     C_R =-m_2^{(i)} V_{ij}^*,\end{array}\right.\\
%  & G^+ \bar{\nu_i} e_j :  &
%  &\left\{\begin{array}{l}C_L = 0,\\
%      C_R = m_j \delta_{ij},\end{array}\right.
%  &&G^- \bar{e_j} \nu_i :     &
%  &\left\{\begin{array}{l}C_L = -m_j \delta_{ij},\\
%      C_R = 0,\end{array}\right.\\
%  & G^{(n)+} \bar{\nu_i} {\mathcal L}^{(n)}_j :  &
%  &\left\{\begin{array}{l}C_L = 0,\\
%      C_R = m^{(j)}_1 \delta_{ij},\end{array}\right.
%  &&G^{(n)-}{\overline{\mathcal L}^{(n)}_j} \nu_i :     &
%  &\left\{\begin{array}{l}C_L = -m^{(j)}_1 \delta_{ij},\\
%      C_R = 0,\end{array}\right.\\
%  & G^{(n)+} \bar{\nu_i} {\mathcal E}^{(n)}_j :  &
%  &\left\{\begin{array}{l}C_L = 0,\\
%      C_R =-m^{(j)}_2 \delta_{ij},\end{array}\right.
%  &&G^{(n)-}{\overline{\mathcal E}^{(n)}_j} \nu_i :     &
%  &\left\{\begin{array}{l}C_L = m^{(j)}_2 \delta_{ij},\\
%      C_R = 0,\end{array}\right.\\
  & H^{(n)+}{\overline{T}^{1(n)}_i} d_j :  &
  &\left\{\begin{array}{l}C_L = -m_3^{(i)} V_{ij},\\
      C_R = M_3^{(i,j)} V_{ij},\end{array}\right.
  &&H^{(n)-}\bar{d_j}T^{1(n)}_i :   &
  &\left\{\begin{array}{l}C_L = -M_3^{(i,j)} V_{ij}^*,\\
     C_R = m_3^{(i)} V_{ij}^*,\end{array}\right.\\
  & H^{(n)+}{\overline{T}^{2(n)}_i} d_j :  &
  &\left\{\begin{array}{l}C_L = m_4^{(i)} V_{ij},\\
      C_R =-M_4^{(i,j)} V_{ij},\end{array}\right.
  &&H^{(n)-}\bar{d_j}T^{2(n)}_i :   &
  &\left\{\begin{array}{l}C_L = M_4^{(i,j)} V_{ij}^*,\\
     C_R =-m_4^{(i)} V_{ij}^*.\end{array}\right.
%  & H^{(n)+} \bar{\nu_i} {\mathcal L}^{(n)}_j :  &
%  &\left\{\begin{array}{l}C_L = 0,\\
%      C_R = m^{(j)}_3 \delta_{ij},\end{array}\right.
%  &&H^{(n)-}{\overline{\mathcal L}^{(n)}_j} \nu_i :     &
%  &\left\{\begin{array}{l}C_L = -m^{(j)}_3 \delta_{ij},\\
%      C_R = 0,\end{array}\right.\\
%  & H^{(n)+} \bar{\nu_i} {\mathcal E}^{(n)}_j :  &
%  &\left\{\begin{array}{l}C_L = 0,\\
%      C_R =-m^{(j)}_4 \delta_{ij},\end{array}\right.
%  &&H^{(n)-}{\overline{\mathcal E}^{(n)}_j} \nu_i :     &
%  &\left\{\begin{array}{l}C_L = m^{(j)}_4 \delta_{ij},\\
%      C_R = 0,\end{array}\right.\\
\end{aligned}
\end{equation}

%\newpage

7) $W^{\mu\pm}{\overline{f}_1}f_2$
  $\displaystyle  :  \frac{i g_2}{\sqrt{2}} \gamma_\mu P_L C_L$, where $C_L$ is given in the following:

\begin{equation}
\begin{aligned}
  & W^{\mu+}\bar{u_i} d_j : &&     C_L = V_{ij},
  && W^{\mu-}\bar{d_j} u_i : &&    C_L = V^*_{ij},\\
  & W^{\mu(n)+}{\overline{T}^{1(n)}_i}d_j : &&   C_L = I^n_1\;c_{in} V_{ij},
  &&W^{\mu(n)-}\bar{d_j}{{T}^{1(n)}_i} : && C_L = I^n_1\;c_{in} V^*_{ij},\\
  & W^{\mu(n)+}{\overline{T}^{2(n)}_i}d_j : &&   C_L = -I^n_1\;s_{in} V_{ij},
  &&W^{\mu(n)-}\bar{d_j}{{T}^{2(n)}_i} : && C_L = -I^n_1\;s_{in}V^*_{ij},
%  & W^{\mu+}\bar {\nu_i} e_j : &&     C_L = \delta_{ij},
%  && W^{\mu-}\bar{e_j} \nu_i : &&    C_L = \delta_{ij},\\
%  & W^{\mu(n)+}\bar {\nu_i} {\mathcal L}^{(n)}_j : &&     C_L = I^n_1\;\delta_{ij},
%  &\hspace{12ex} & W^{\mu(n)-}{\overline{\mathcal L}^{(n)}_j} \nu_i : &&    C_L =  I^n_1\;\delta_{ij},\\
%  & W^{\mu(n)+}\bar {\nu_i} {\mathcal E}^{(n)}_j : &&     C_L = 0,
%  && W^{\mu(n)-}{\overline{\mathcal E}^{(n)}_j} \nu_i : &&    C_L = 0.
\end{aligned}
\end{equation}
where the fermion fields $f\equiv u, d, T^1_t, T^2_t$.

The mass parameters $m_x^{(i)}$ are given in the following \cite{Datta:2015aka}:
\begin{equation}
\label{mparameters}
  \begin{aligned}
    m_1^{(i)} &= I^n_2\;m_{V^{(n)}}c_{in} +I^n_1\;m_i s_{in},\\
    m_2^{(i)} &= -I^n_2\;m_{V^{(n)}}s_{in}+I^n_1\;m_i c_{in},\\
    m_3^{(i)} &= -I^n_2\;iM_W c_{in} +I^n_1\;i\frac{m_{V^{(n)}}m_i}{M_W}s_{in},\\
    m_4^{(i)} &= I^n_2\;iM_W s_{in}+I^n_1\;i\frac{m_{V^{(n)}}m_i}{M_W}c_{in},
  \end{aligned}
\end{equation}
where $m_i$ denotes the mass of the zero-mode {\it up-type} fermion and $c_{in}=\cos(\alpha_{in})$ and $s_{in}=\sin(\alpha_{in})$ with $\alpha_{in}$ as defined earlier.

And the mass parameters $M_x^{(i,j)}$ are given in the following \cite{Datta:2015aka}:
\begin{equation}\label{Mparameters}
  \begin{aligned}
    M_1^{(i,j)}  &= I^n_1\;m_j c_{in},\\
    M_2^{(i,j)}  &= I^n_1\;m_j s_{in},\\
    M_3^{(i,j)}  &= I^n_1\;i\frac{m_{V^{(n)}}m_j}{M_W}c_{in},\\
    M_4^{(i,j)}  &= I^n_1\;i\frac{m_{V^{(n)}}m_j}{M_W}s_{in},
  \end{aligned}
\end{equation}
where $m_j$ denotes the mass of the zero-mode {\it down-type} fermion. 

In all the Feynman vertices the factors $I^n_1$ and $I^n_2$ are represented as the overlap integrals given in the following \cite{Datta:2015aka}

\begin{equation}
I^n_1 = 2\sqrt{\frac{1+\frac{r_V}{\pi R}}{1+\frac{r_f}{\pi R}}}\left[ \frac{1}{\sqrt{1 + \frac{r^2_f m^2_{f^{(n)}}}{4} + \frac{r_f}{\pi R}}}\right]\left[ \frac{1}{\sqrt{1 + \frac{r^2_V m^2_{V^{(n)}}}{4} + \frac{r_V}{\pi R}}}\right]\frac{m^2_{V^{(n)}}}{\left(m^2_{V^{(n)}} - m^2_{f^{(n)}}\right)}\frac{\left(r_{f} - r_{V}\right)}{\pi R},
\label{i1}
\end{equation}

\begin{equation}
I^n_2 = 2\sqrt{\frac{1+\frac{r_V}{\pi R}}{1+\frac{r_f}{\pi R}}}\left[ \frac{1}{\sqrt{1 + \frac{r^2_f m^2_{f^{(n)}}}{4} + \frac{r_f}{\pi R}}}\right]\left[ \frac{1}{\sqrt{1 + \frac{r^2_V m^2_{V^{(n)}}}{4} + \frac{r_V}{\pi R}}}\right]\frac{m_{V^{(n)}}m_{f^{(n)}}}{\left(m^2_{V^{(n)}} - m^2_{f^{(n)}}\right)}\frac{\left(r_{f} - r_{V}\right)}{\pi R}.
\label{i2}
\end{equation}
\end{appendices}

%\bibliographystyle{jhep}
%\bibliography{btoxsll}

\providecommand{\href}[2]{#2}\begingroup\raggedright\endgroup

\end{document}